\newcommand{\be}{\begin{equation}}
\newcommand{\ee}{\end{equation}}
\newcommand{\bea}{\begin{eqnarray}}
\newcommand{\eea}{\end{eqnarray}}
\newcommand{\ba}[1]{\begin{array}{#1}}
\newcommand{\ea}{\end{array}}

\newcommand{\bt}{\begin{tabular}}
\newcommand{\et}{\end{tabular}}
\newcommand{\Tr}{{\rm Tr}}

\newcommand{\pa}{\partial}
\newcommand{\beas}{\begin{eqnarray*}}
\newcommand{\eeas}{\end{eqnarray*}}

\newcommand{\fr}{\frac}

\newcommand{\dg}{\dagger}

\newcommand{\pam}{\partial_\mu}

\documentclass[amsmath,12pt,amssymb,preprint,nofootinbib]{revtex4}
\usepackage{amsfonts} 
\usepackage{graphicx} 
\usepackage{epsfig}
\usepackage{multirow}
\usepackage{bm}
\usepackage{appendix}

\begin{document}

\title
{\textbf {Dirac sea effects on Heavy Quarkonia decay widths 
in magnetized matter 
-- a field theoretical model of composite hadrons}}
\author{Amruta Mishra}
\email{amruta@physics.iitd.ac.in}
\affiliation{Department of Physics, IIT, Delhi,
New Delhi -- 110 016, India}
\author{S.P. Misra}
\email{misrasibaprasad@gmail.com}
\affiliation{Institute of Physics, Bhubaneswar -- 751005, India}
\begin{abstract}
We study the partial decay widths of charmonium
(bottomonium) states to ${\rm D\bar D \; (B\bar B)}$ mesons
in magnetized (nuclear) matter using a field theoretical model 
of composite hadrons with quark (and antiquark) constituents. 
These are computed from the mass modifications 
of the decaying and produced mesons 
within a chiral effective model,
including the nucleon Dirac sea effects.
The mass modifications of the open charm (bottom) mesons 
are calculated from their interactions with the nucleons 
and the scalar mesons, 
whereas the mass shift of the heavy quarkonium state is obtained 
from the medium change of a scalar dilaton field, $\chi$, 
which mimics the gluon condensates of QCD.
The Dirac sea contributions are observed 
to lead to a rise (drop) in the quark condensates
as the magnetic field is increased, 
an effect called the (inverse) magnetic catalysis. 
These effects are observed to be significant 
and the anomalous magnetic moments (AMMs) of the nucleons are observed
to play an important role. For $\rho_B$=0, there is observed to be
magnetic catalysis (MC) without and with AMMs, whereas,   
for $\rho_B=\rho_0$, the inverse magnetic catalysis (IMC) is observed 
when the AMMs are taken into account, contrary to MC,
when the AMMs are ignored. In the presence of a magnetic field,
there are also mixings of spin 0 (pseudoscalar) and spin 1
(vector) states (PV mixing) which modify the masses of these mesons.
The magnetic field effects on the heavy quarkonium decay widths
should have observable consequences on the production 
the heavy flavour mesons, which are created in the early stage
of ultra-relativistic peripheral heavy ion collisions, at RHIC and LHC, 
when the produced magnetic fields can still be extremely large.
\end{abstract}
\maketitle

\def\bfm#1{\mbox{\boldmath $#1$}}
\def\bfs#1{\mbox{\bf #1}}

\vspace{-1cm}
\section{Introduction}
The study of the in-medium properties of the heavy flavour mesons
\cite{Hosaka}, in particular in the presence of strong magnetic fields, 
has been a topic of intense research due to its relevance
in relativistic heavy ion collision experiments. 
The heavy flavour mesons are created at the early stage 
when the magnetic fields resulting from ultra-relativistic 
peripheral heavy ion collisions, estimated to be huge \cite{tuchin}, 
can still be extremely large.
The heavy quarkonium states and the open heavy flavour mesons
have been studied extensively in the literature 
using the potential models 
\cite{eichten_1,eichten_2,satz_1,satz_2,satz_3,satz_4,satz_5,repko,
Ebert,Bonati_pot_model,Yoshida_Suzuki_heavy_flavour_meson_strong_B},
the QCD sum rule approach
\cite{kimlee,klingl,amarvjpsi_qsr,jpsi_etac_mag,upsilon_etab_mag,moritalee_1,moritalee_2,moritalee_3,moritalee_4,open_heavy_flavour_qsr_1,open_heavy_flavour_qsr_2,open_heavy_flavour_qsr_3,open_heavy_flavour_qsr_4,Wang_heavy_mesons_1,Wang_heavy_mesons_2,arvind_heavy_mesons_QSR_1,arvind_heavy_mesons_QSR_2,arvind_heavy_mesons_QSR_3},
the coupled channel approach
\cite{ltolos,ljhs,mizutani_1,mizutani_2,HL,tolos_heavy_mesons_1,tolos_heavy_mesons_2}, the quark meson coupling (QMC) model
\cite{open_heavy_flavour_qmc_1,open_heavy_flavour_qmc_2,open_heavy_flavour_qmc_3,qmc_1,qmc_2,qmc_3,qmc_4,krein_jpsi,krein_17},
as well as using a chiral effective model
\cite{amdmeson,amarindamprc,amarvdmesonTprc,amarvepja,DP_AM_Ds,DP_AM_bbar,DP_AM_Bs,AM_DP_upsilon}.
Studies of heavy quarkonium states ($\bar Q Q$ bound states, $Q=c,b$) 
in presence of a gluon field, assuming the distance between 
$Q$ and $\bar Q$ to be small as compared to the scale of the 
gluonic fluctuations, show that the mass modifications of these states
arise from the medium modification
of the scalar gluon condensate in the leading order 
\cite{pes1,pes2,voloshin}.
A study of the mass modifications of the charmonium states  
due to the gluon condensates as well as
$\bar D D$ meson loop \cite{leeko} showed that
the dominant contributions are due to the medium
modifications of the gluon condensates.
In a chiral effective model, 
the in-medium masses of the heavy quarkonium (charmonium 
and bottomonium) have been computed 
from the medium change of a scalar dilaton field
\cite{amarvdmesonTprc,amarvepja,AM_DP_upsilon},
which simulates the gluon condensates of QCD within
the effective hadronic model.

The chiral effective model, in the original version
with three flavours of quarks (SU(3) model) 
\cite{Schechter,paper3,kristof1,hartree},
has been used extensively in the literature,
for the study of finite nuclei \cite{paper3},
strange hadronic matter \cite{kristof1},
light vector mesons \cite{hartree},
strange pseudoscalar mesons, e.g. the kaons and antikaons
\cite{kaon_antikaon,isoamss,isoamss1,isoamss2}
in isospin asymmetric hadronic matter,
as well as for the study of bulk matter of neutron stars
\cite{pneutronstar}.
Within the QCD sum rule framework,
the light vector mesons
\cite{am_vecmeson_qsr,vecqsr_mag},
as well as, the heavy quarkonium states
\cite{amarvjpsi_qsr,jpsi_etac_mag,upsilon_etab_mag},
in (magnetized) hadronic matter have been
studied, using the medium changes of the light quark condensates
and gluon condensates calculated within the
chiral SU(3) model. 
Using the in-medium masses of the heavy flavour mesons in the 
(magnetized) hadronic matter,
calculated within the chiral effective model, the partial decay widths
of the heavy quarkonium states to the open heavy flavour mesons
have been studied in (magnetized) hadronic medium
\cite{amarvepja,charmdecay_mag}, using a light quark-antiquark 
pair creation model \cite{friman},
namely the $^3P_0$ model \cite{3p0_1,3p0_2,3p0_3,3p0_4} as well as
using a field theoretical model for composite hadrons
with quark (and antiquark) constituents 
\cite{amspmwg,charmdw_mag,open_charm_mag_AM_SPM,
amspm_upsilon,upslndw_mag}.
The effects of magnetic field on the masses of the heavy flavour mesons
have been studied in Refs.
\cite{B_mag_QSR,machado_1,Alford_Strickland_2013,charmonium_mag_QSR,charmonium_mag_lee,Suzuki_Lee_2017,Quarkonia_B_Iwasaki_Oka_Suzuki,Gubler_D_mag_QSR},
and, it is observed that the spin-magnetic field interaction leads to 
mixing between the pseudoscalar meson and the longitudinal
component of the vector meson (PV mixing). This
results in a dominant rise (drop) in the mass of the longitudinal 
component of the vector meson (pseudoscalar) meson
for the heavy quarkonia (charmonia and bottomonia) states
as well as for open charm (bottom) mesons 
\cite{charmonium_mag_QSR,charmonium_mag_lee,Suzuki_Lee_2017,Alford_Strickland_2013,Quarkonia_B_Iwasaki_Oka_Suzuki,Gubler_D_mag_QSR}. 
In the presence of a magnetic field, the studies of the effects 
of Dirac sea (DS) in the quark matter sector 
\cite{kharzeevmc,kharmc1,elia,chernodub} 
within the Nambu-Jona-Lasinio model 
\cite{Preis,menezes,ammc}, are observed to lead to enhancement
of the light quark condensates with increase in the magnetic field,
an effect called the magnetic catalysis (MC). 
The opposite trend of the light quark condensates
with magentic field, namely the inverse magnetic catalysis 
(IMC) is observed in some lattice QCD calculations \cite{balicm}, 
where the crittical temperature, $T_c$ is seen to decrease 
with increase in the magnetic field.
For the nuclear matter, the effects of Dirac sea (DS) have 
been studied using the Walecka model as well as an extended 
linear sigma model in Ref. \cite{haber}. These are observed to
lead to magnetic catalysis (MC) effect for zero temperature
and zero density, which is observed as a rise 
in the effective nucleon mass with the increase in magnetic field.
In Ref. \cite{arghya}, the contributions of Dirac sea
of the nucleons to the self-energies of the nucleons
have been studied in the Walecka model 
by summing over the scalar ($\sigma$) and vector ($\omega$) 
tadpole diagrams, in a weak magnetic field 
approximation of the fermion propagator. 
At zero density, the effects of the Dirac sea
are seen to lead to magnetic catalysis
(MC) effect at zero temperature \cite{arghya}. 
When the anomalous magnetic moments (AMMs)
of the nucleons are taken into account,
at a finite density and zero temperature, 
there is observed to be a drop in the effective nucleon 
mass with increase in the magnetic field. This behaviour with
the magnetic field is observed when the temperature
is raised from zero to non-zero values, upto the critical 
temperature, $T_c$, when the nucleon mass has a sudden 
drop, corresponding to the vacuum to nuclear matter
phase transition. The  decrease in $T_c$ with increase in
value of $B$ is identified with the inverse magnetic catalysis
(IMC) \cite{arghya}. 

In the present work, the partial decay widths of the charmonium
(bottomonium) states to open heavy flavour mesons, $D\bar D (B\bar B$) 
are studied in magnetized (nuclear) matter 
using a field theoretical model of composite hadrons.
As the matter created in ultra-relativiistic peripheral heavy ion
collisions is dilute, we study the partial decay widths of
the lowest quarkonium states in the charm and bottom sectors,
$\psi(3770)$ and $\Upsilon(4S)$ (which decay to $D\bar D$ and 
$B\bar B$ in vacuum). These are investigated for $\rho_B=0$ 
as well as for $\rho_B=\rho_0$, the nuclear matter saturation density, 
for symmetric as well as asymmetric nuclear matter in the 
presence of an external magnetic field.
The study of effects of temperature on the open charm and charmonium
masses (and hence on the charmonium decay widths) 
\cite{amarvdmesonTprc,amarvepja}
have been observed to be marginal for small densities
(upto $\rho_0$). Within the chiral effective model, 
the mass shift of the heavy quarkonium states and 
the open heavy flavour mesons arise from the
medium modifications of the dilaton field and the scalar fields,
which have marginal modifications due to temperature, and, hence
the temperature effects on the quarkonium decay widths 
(due to mass modification of these mesons) are not taken 
into account in the present study. 
The magnetic effects are the most dominant effects for the
(dilute) matter resulting from ultra-relativistic peripheral collisons,
which include the contributions from the magnetized Dirac sea of nucleons
as well as PV mixing, in additon to the Landau level contributions
for the charged hadrons.
In the chiral effective model, the effects of the Dirac sea are incorporated
to the nucleon propagator, through summation of scalar 
($\sigma$, $\zeta$ and $\delta$) and vector ($\omega$ and $\rho$) 
tadpole diagrams. 
When the anomalous 
magnetic moments (AMMs) of the nucleons are not taken into 
account, for zero density as well as for $\rho_B=\rho_0$, 
magnetic catalysis (MC) is observed. However, when the AMMs 
of nucleons are considered, for $\rho_B=\rho_0$
(both for symmetric and asymmetric nuclear matter), 
inverse magnetic catalysis (IMC)
is observed, i.e., the quark condensate is observed to be reduced
with rise in the magnetic field.

The outline of the paper is as follows.
In section II, we describe briefly the chiral effective model
used to calculate the masses of the charmonium (bottomonium) 
and open charm (bottom) mesons,
accounting for the effects of the Dirac sea for the nucleons.
The PV mixing effects are also taken into account which modify the masses
of the heavy quarkonium states as well as open heavy flavour mesons.
In section III, the computations of the decay widths
of $\psi(3770)\rightarrow D\bar D$ and $\Upsilon(4S)\rightarrow
B\bar B$ using the field theoretical model
of composite hadrons are briefly described, and,
the salient features of the model are presented in
Appendix A. The results of the
partial decay widths in magnetized (nuclear)
matter are discussed in section IV and the summary
of the present work are given in section V.

\section{Mass modifications of charm and bottom mesons}
We describe breifly the chiral effective model used to study
the open charm (bottom) mesons and the charmonium (bottomonium)
states in magnetized nuclear matter.
The model is a generalization of a chiral SU(3) model,
based on a nonlinear realization of chiral 
symmetry, and, the breaking of scale invariance of QCD.
The scale symmetry breaking is incorporated through a scalar
dilaton field (which mimics the scalar gluon condensate) 
and the mass modifications of the heavy quarkonium
states are obtained from medium modifications of the dilaton field.
The in-medium masses of the open heavy (charm and bottom) 
flavour mesons are obtained by generalizing the chiral SU(3) model
to include the interactions of the open charm and bottom mesons
with the light hadrons.
 
In the presence of a magnetic field, the Lagrangian for SU(3) model 
has the form \cite{kmeson_mag}
\begin{equation}
{\cal L} = {\cal L}_{kin} + \sum_ W {\cal L}_{BW}
          +  {\cal L}_{vec} + {\cal L}_0
+ {\cal L}_{scalebreak}+ {\cal L}_{SB}+{\cal L}_{mag}^{B\gamma},
\label{genlag} \end{equation}
where ${\cal L}_{kin}$ refers to the kinetic energy terms
of the baryons and the mesons, 
${\cal L}_{BW}$ is the baryon-meson interaction term,
${\cal L}_{vec}$  describes the dynamical mass generation of the vector 
mesons via couplings to the scalar mesons and contain additionally 
quartic self-interactions of the vector fields, ${\cal L}_{0}$ contains 
the meson-meson interaction terms,
${\cal L}_{scalebreak}$ is the scale invariance breaking 
term and ${\cal L}_{SB}$ describes the explicit 
chiral symmetry breaking. 
The kinetic energy terms are given as
\begin{eqnarray}
\label{kinetic}
{\cal L}_{kin} &=& i{\rm Tr} \overline{B} \gamma_{\mu} D^{\mu}B
                + \frac{1}{2} {\rm Tr} D_{\mu} X D^{\mu} X
+  {\rm Tr} (u_{\mu} X u^{\mu}X +X u_{\mu} u^{\mu} X)
                + \frac{1}{2}{\rm Tr} D_{\mu} Y D^{\mu} Y 
 \nonumber \\
               &+&\frac {1}{2} D_{\mu} \chi D^{\mu} \chi
                - \frac{ 1 }{ 4 } {\rm Tr}
\left(\tilde V_{ \mu \nu } \tilde V^{\mu \nu }  \right)
- \frac{ 1 }{ 4 }{\rm  Tr} \left( {\cal A}_{ \mu \nu } {\cal A}^{\mu \nu }
 \right)
- \frac{ 1 }{ 4 }{\rm  Tr} \left(F_{ \mu \nu } F^{\mu \nu }  \right),
\end{eqnarray}
where, $B$ is the baryon octet, $X$ is the scalar meson
multiplet, $Y$ is the pseudoscalar chiral singlet, 
$\chi$ is the scalar dilaton field,
${V}_{\mu\nu}=\pa_\mu{V}_\nu-\pa_\nu{V}_\mu$,
${\cal A}_{\mu\nu}= \pa_\mu{\cal A}_\nu-\pa_\nu{\cal A}_\mu $, 
and $F_{\mu\nu}= \pa_\mu A_\nu-\pa_\nu A_\mu $, 
are the field strength tensors of 
the vector meson multiplet, ${V}^\mu$, 
the axial vector meson multiplet ${\cal A}^\mu$ 
and the photon field, $A^\mu$.
In Eq. (\ref{kinetic}), 
\begin{equation}
u_\mu= -\fr{i}{4} [(u^\dg(\pam u)-(\pam u^\dg) u) 
 - (u (\pam u^\dg)-(\pam u) u^\dg)],
\label{u_mu}
\end{equation}
and the covariant derivative of a field $\Phi (\equiv B,X,Y,\chi)$
reads $ D_\mu \Phi = \pam\Phi + [\Gamma_\mu,\Phi]$, with
\begin{equation}
\Gamma_\mu=-\fr{i}{4} [(u^\dg(\pam u)-(\pam u^\dg) u)
 + (u (\pam u^\dg)-(\pam u) u^\dg)],
\label{gamma_mu}
\end{equation}
where $u=\exp\big[\fr{i}{\sigma_0}\pi^a\lambda^a\gamma_5\big]$,
with $\pi^a$ and $\lambda^a$, $i=1,..8$, as the pseudoscalar mesons
and the Gell-Mann matrices.
The interaction of the baryons with the meson, $W$ (scalar,
pesudoscalar, vector, axialvector meson) 
is given as 
\begin{equation}
{\cal L}_{BW}=-\sqrt 2 g_8^W\Big ( \alpha_W [\bar B O B W]_F
+ (1- \alpha_W) [\bar B O B W]_D\Big )-\frac {g_1^W}{\sqrt 3}
Tr(\bar B O B W)tr (W),
\label{baryon_meson_int}
\end{equation}
where, the $F$-type (antisymmetric) and $D$-type (symmetric)
couplings are defined as
$[\bar B O B W]_F=Tr(\bar B O W B-\bar B O B W)$ and
$ [\bar B O B W]_D=Tr(\bar B O W B+\bar B O B W)
-\frac{2}{3} Tr(\bar B O B)Tr(W)$.
In equation (\ref{baryon_meson_int}), 
$(W,O)\equiv (X,1), (u,\gamma^5),  (V,\gamma^\mu)\;\; {\rm and}\;\; 
({\cal A},\gamma^\mu \gamma^5)$, for the interactions of the
baryons with the scalar, the pseudoscalar, the vector and
the axial-vector mesons respectively.

The Lagrangian for the vector meson interaction is written as
\bea
{\cal L}_{vec} &=&
    \fr{m_V^2}{2}\fr{\chi^2}{\chi_0^2}\Tr\big({V}_\mu{V}^\mu\big)
+   \fr{\mu}{4}\Tr\big({V}_{\mu\nu}{V}^{\mu\nu}X^2\big) 
+ \fr{\lambda_V}{12}\Big(\Tr\big({V}^{\mu\nu}\big)\Big)^2 +
    2({g}_4)^4\Tr\big({V}_\mu{V}^\mu\big)^2  \, .
\eea
The masses of $\omega,\rho$ and $\phi$ are fitted from $m_V, \mu$ and
$\lambda_V$.
The Lagrangian describing the interaction for the scalar mesons, $X$,
and pseudoscalar singlet, $Y$, is given as \cite{paper3}
\bea
\label{cpot}
{\cal L}_0 &= &  -\frac{ 1 }{ 2 } k_0 \chi^2 I_2
     + k_1 (I_2)^2 + k_2 I_4 +2 k_3 \chi I_3,
\eea with $I_2= \Tr (X+iY)^2$, $I_3=\det (X+iY)$ and $I_4 = \Tr
(X+iY)^4$. 
In the above, $\chi$ is the scalar dilaton
field which is introduced in order to mimic the QCD trace
anomaly, i.e. the non-vanishing energy-momentum tensor
\be
\theta_\mu^\mu = (\beta_{QCD}/2g)\langle
G_{\mu\nu}^{a} G^{\mu\nu a} \rangle + \sum_i m_i \bar {q_i} q_i,
\label{tensorquark}
\ee
where $G^a_{\mu\nu}$ is the gluon field tensor,
and, the second term in the trace accounts for the finite 
quark masses, with $m_i$ as the current quark mass for the quark
of flavor, $i=u,d,s$. 
The scale breaking 
and the explicit chiral symmetry breaking terms are given as 
\cite{Schechter,paper3}
\bea
\label{lscale}
 && {\cal L}_{\mathrm{scalebreak}}=- \frac{1}{4}\chi^4 \ln
   \frac{ \chi^4 }{ \chi_0^4}
 +\frac{d}{3}\chi^4 \ln \Bigg( \Big(\frac{I_3}{\det \langle X \rangle_0}
\Big) \Big (\frac{\chi}{\chi_0}\Big)^3 \Bigg),\\
\label{scalebreak}
&& {\cal L}_{SB}=\Tr A_p\left(u(X+iY)u+u^\dagger(X-iY)u^\dagger\right),
\label{esb}
\eea
with $A_p=1/\sqrt{2}m_\pi^2 f_\pi{\mathrm{diag}}(1,1,
\frac{2 m_K^2 f_K}{m_{\pi}^2 f_\pi}-1)$,
here $m_{\pi}$ and $m_K$ are the masses of the pion and K-meson, 
and, $f_\pi$ and $f_K$, their decay widths.

In the present investigation, we use the mean field approximation,
where all the meson fields are treated as classical fields. 
In this approximation, only the scalar and the vector fields 
contribute to the baryon-meson interaction, ${\cal L}_{BW}$
since for all the other mesons, the expectation values are zero.
The various terms of the Lagrangian density in the mean field 
approximation are given as
\begin{eqnarray}
{\cal L}_{BX}+{\cal L}_{BV} &=& -\sum_i\overline{\psi_{i}}\, [g_{i
\omega}\gamma_0 \omega + g_{i\phi}\gamma_0 \phi
+m_i^{\ast} ]\,\psi_{i} 
\label{lagBW}\\
{\cal L}_{vec} &=& \frac{1}{2} \frac{\chi^2}{\chi_0^2}\Big(
m_{\omega}^{2} \omega^ 2+m_{\rho}^{2} \omega^ 2+m_{\phi}^{2} \omega^ 2
\Big) +g_4^4 (\omega^4 +2 \phi^4+6 \omega^2 \rho^2+\rho^4)\\
{\cal L}_{0} &=& - \frac{ 1 }{ 2 } k_0 \chi^2
(\sigma^2+\zeta^2+\delta^2) + k_1 (\sigma^2+\zeta^2+\delta^2)^2
\nonumber \\
     &+& k_2 ( \frac{ \sigma^4}{ 2 } + \frac{\delta^4}{2} + \zeta^4)
     + k_3 \chi (\sigma^2 - \delta^2) \zeta - k_4 \chi^4 \\
{\cal L}_{scalebreak}&= &-\frac{1}{4} \chi^{4} 
{\rm ln} \frac{\chi^{4}}{\chi_{0}^{4}} + \frac{d}{3} \chi^{4} 
{\rm ln} \Bigg( \frac{\left( \sigma^{2} - \delta^{2}\right)\zeta }
{\sigma_{0}^{2} \zeta_{0}} \Big( \frac{\chi}{\chi_{0}}\Big) ^{3}\Bigg), 
\label{scalebreak}
\end{eqnarray}
The baryon-scalar meson interactions generate the baryon masses
and the parameters corresponding to these interactions are adjusted
so as to obtain the baryon masses as their experimentally measured
vacuum values. In equation (\ref{lagBW}), the effective mass of the baryon 
of type $i$ ($i =p,n,\Lambda,\Sigma^{\pm,0},\Xi^{0,-}$) is given as 
\be
m_i^* = -g_{\sigma i}{\sigma}-g_{\zeta i}{\zeta}
-g_{\delta i}{\delta},  
\label{m_B_eff}
\ee
which is calculated from the values of the scalar fields
in the magnetized medium, and, the masses with the vacuum values
of the scalar fields correspond to the experimentally measured 
vacuum values of the baryons.   

The explicit chiral symmetry breaking term is given as 
\begin{eqnarray}
{\cal L} _{SB} & = & {\rm Tr} \left [ {\rm diag} \left (
-\frac{1}{2} m_{\pi}^{2} f_{\pi} (\sigma+\delta), 
-\frac{1}{2} m_{\pi}^{2} f_{\pi} (\sigma-\delta), 
\Big( \sqrt{2} m_{k}^{2}f_{k} - \frac{1}{\sqrt{2}} 
m_{\pi}^{2} f_{\pi} \Big) \zeta \right) \right ]\nonumber \\
\label{ecsb}
&=& -\left[m_{\pi}^2 f_{\pi} \sigma
+ (\sqrt{2}m_K^2 f_K - \frac{ 1 }{ \sqrt{2} } m_{\pi}^2 f_{\pi})\zeta
\right].
\label{lag_SB_MFT}
\end{eqnarray}
In the above, the matrix, whose trace gives the Lagrangian density 
corresponding to the explicit chiral symmetry breaking in the chiral 
SU(3) model, has been explicitly written down. 
Comparing the above term with the explicit chiral symmetry 
breaking term of the Lagrangian density in QCD given as
\begin{eqnarray}
{\cal L}^{QCD}_{SB} & =- & {\rm Tr} \left [ {\rm diag} \left (m_u \bar u u, 
m_d \bar d d , m_s \bar s s \right ) \right],
\label{ecsbqcd}
\end{eqnarray}
one obtains the nonstrange quark condensates ($\langle \bar u u \rangle$ and 
$\langle \bar d d \rangle$) and the strange quark condensate 
($\langle \bar s s \rangle $)  to be related to the
the scalar fields, $\sigma$, $\delta$ and $\zeta$
as 
\begin{eqnarray}
m_u\langle \bar u u \rangle 
&=& \frac{1}{2}m_{\pi}^{2} f_{\pi} (\sigma+\delta);\;\;\;
m_d \langle \bar d d \rangle
= \frac{1}{2}m_{\pi}^{2} f_{\pi} (\sigma-\delta); \nonumber\\
&& m_s\langle \bar s s \rangle 
= \Big( \sqrt {2} m_{k}^{2}f_{k} - \frac {1}{\sqrt {2}} 
m_{\pi}^{2} f_{\pi} \Big) \zeta.
\label{qbarq_scalar}
\end{eqnarray}
It might be noted here that with the choice for $A_p$ in the explicit 
symmetry breaking term as given by equation (\ref{esb}), 
together with the constraints
$\sigma_0=-f_\pi$, $\zeta_0=-\frac {1}{\sqrt 2} (2 f_K -f_\pi)$
assure that the PCAC-relations of the pion and kaon are fulfilled.
Using one loop QCD $\beta$ function 
$\beta_{\rm {QCD}} \left( g \right) = -\frac{11 N_{c} g^{3}}{48 \pi^{2}} 
\left( 1 - \frac{2 N_{f}}{11 N_{c}} \right)$,
with $N_c=3$, the number of colors and $N_f$ as the number of
quark flavor,
in the trace of energy momentum tensor in QCD given by equation 
(\ref{tensorquark}) and equating with $\theta^\mu_\mu$ of the
chiral model 
\be
\theta_\mu^\mu
= \chi \frac{\partial {\cal L}}{\partial \chi} - 4{\cal L} 
=(1-d)\chi^4,
\label{tensor1}
\ee
the scalar gluon condensate gets related to the dilaton fleld as 
\cite{amarvepja}
\begin{equation}
\langle \frac{\alpha_{s}}{\pi} 
G_{\mu\nu}^{a} G^{\mu\nu a} \rangle 
= \frac{24}{(33-2N_f)}(1-d)\chi^{4}. 
\label{chiglu}
\end{equation} 
in the limiting situation of massless quarks
in the energy momentum tensor of QCD given by equation 
(\ref{tensorquark}). 

The term ${\cal L}_{mag}^{B\gamma}$ in the Lagrangian 
given by equation (\ref{genlag}), describes the interacion
of the baryons with the electromagnetic field, and,
is given as 
\cite{dmeson_mag,bmeson_mag,charmonium_mag}
\be 
{\cal L}_{mag}^{B\gamma}=-{\bar {\psi_i}}q_i 
\gamma_\mu A^\mu \psi_i
-\frac {1}{4} \kappa_i \mu_N {\bar {\psi_i}} \sigma ^{\mu \nu}F_{\mu \nu}
\psi_i,
\label{lmag_Bgamma}
\ee
where, $\psi_i$ corresponds to the $i$-th baryon.
The tensorial interaction of baryons 
with the electromagnetic field given by the second term 
in the above equation is related to the
anomalous magnetic moments of the baryons.
We choose the magnetic field to be uniform and along the
z-axis, and take the vector potential to be
$A^\mu =(0,0,Bx,0)$. The number and scalar densities 
of the proton have contributions from the Landau energy levels
and the neutrons have contributions to their number and scalar densities
due to the anomalous magnetic moment, in the presence
of a magnetic field \cite{dmeson_mag,bmeson_mag}. 
The expresssions for the number and scalar densities of the proton
in the presence of a uniform magnetic field (chosen to be
along z-direction) and accounting for the anomalous magnetic moments
for the nucleons are given as 
\cite{broderick1,broderick2,Wei}
\begin{equation}
\rho_p=\frac{eB}{4\pi^2} \Bigg [ 
\sum_{\nu=0}^{\nu_{max}^{(S=1)}} k_{f,\nu,1}^{(p)} 
+\sum_{\nu=1}^{\nu_{(max)}^{(S=-1)}} k_{f,\nu,-1}^{(p)} 
\Bigg]
\label{rhop_mag}
\end{equation}
and 
\begin{eqnarray}
\rho^s_p &=& \frac{eB{m_p^*}}{2\pi^2} \Bigg [ 
\sum_{\nu=0}^{\nu_{max}^{(S=1)}}
\frac {\sqrt {{m_p^*}^2+2eB\nu}+\Delta_p}{\sqrt {{m_p^*}^2+2eB\nu}}
\ln |\frac{ k_{f,\nu,1}^{(p)} + E_f^{(p)}}{\sqrt {{m_p^*}^2
+2eB\nu}+\Delta_p}|\nonumber \\
&+&\sum_{\nu=1}^{\nu_{max}^{(S=-1)}}
\frac {\sqrt {{m_p^*}^2+2eB\nu}-\Delta_p}{\sqrt {{m_p^*}^2+2eB\nu}}
\ln|\frac{ k_{f,\nu,-1}^{(p)}+E_f^{(p)}}{\sqrt {{m_p^*}^2
+2eB\nu}-\Delta_p}|\Bigg]+ {\rho^s_p}^{(DS)}.
\label{rhps_mag}
\end{eqnarray}
where, $k_{f,\nu,\pm 1}^{(p)}$ are the Fermi momenta of protons
for the Landau level, $\nu$ for the spin index, $S=\pm 1$,
i.e. for spin up and spin down projections for the proton.
These Fermi momenta are related to the Fermi energy of the
proton as
\begin{equation}
k_{f,\nu,S}^{(p)}=\sqrt { {E_f^{(p)}}^2
-\Big (
{\sqrt {{m_p^*}^2+2eB\nu}+S\Delta_p}\Big )^2}.
\label{kfp_mag}
\end{equation}
The number density and the scalar density of neutrons are given as
\begin{eqnarray}
\rho_{n}=  \frac{1}{4\pi^2} \sum _{S=\pm 1}
\Bigg \{ \frac{2}{3} {k_{f,S}^{(n)}}^3
+S\Delta_n \Bigg[ (m_n^*+S\Delta_n) k_{f,S}^{(n)}
+ {E_f^{(n)}}^2 \Bigg(sin^{-1} \Big (
\frac{m_n^*+S\Delta_n}{E_f^{(n)}}\Big)-\frac{\pi}{2}\Bigg)\Bigg]
\Bigg \}
\label{rhon_mag}
\end{eqnarray}
and
\begin{equation}
\rho^s_n =\frac{m_n^*}{4\pi^2} \sum _{S=\pm 1} 
\Bigg [ k_{f,S}^{(n)} E_f^{(n)} - 
(m_n^*+S\Delta_n)^2 \ln | \frac {k_{f,S}^{(n)}+ 
E_f^{(n)}}{m_n^*+S\Delta_n} | \Bigg] + {\rho^s_n}^{(DS)}.
\label{rhns_mag}
\end{equation}
The Fermi momentum, $k_{f,S}^{(n)}$ 
for the neutron with spin projection, S 
($S=\pm 1$ for the up (down) spin projection), 
is related to the Fermi energy for the 
neutron, $E_f^{(n)}$ as
\begin{equation}
k_{f,S}^{(n)}= \sqrt { {E_f^{(n)}}^2 -
(m_n^*+S\Delta_n)^2}.
\label{kfn_mag}
\end{equation}
In the equations (\ref{rhop_mag})-(\ref{kfn_mag}),
the parameter $\Delta _{i}$ is related 
to the anomalous magnetic moment for the nucleon, $i$
($i=p,n$) as $\Delta_i =-\frac{1}{2} \kappa_i \mu_N B$,
where, $\kappa_i$, occurring in the second term
in the Lagrangian density given by Eq. (\ref{lmag_Bgamma}),
is the value of the gyromagnetic ratio of the nucleon
corresponding to the anomalous magnetic moment 
of the nucleon. 
In the present study of magnetized (nuclear) matter, 
the meson fields are treated as classical 
in the mean field approximation,
and nucleons as quantum fields and the self energies of the nucleons
include the contributions from the Dirac sea.
In addition to using the mean field approximation,
where the meson fields are replaced by their expectation
values, we also use the approximations that
$\bar \psi_i \psi_j = \delta_{ij} \langle \bar \psi_i \psi_i
\rangle \equiv \delta_{ij} \rho_i^s $
and $\bar \psi_i \gamma^\mu \psi_j = \delta_{ij} \delta^{\mu 0} 
\langle \bar \psi_i \gamma^ 0 \psi_i
\rangle \equiv \delta_{ij} \delta^{\mu 0} \rho_i $, 
where, $\rho_i^s$ and 
$\rho_i$ are the scalar and number density of fermion of species, 
$i$ (neutron and proton in the present investigation).
Using the scalar densities of the nucleons in the presence
of magnetic field, the values of the scalar fields, 
$\sigma$, $\zeta$ and $\delta$
are obtained by solving their coupled equations of motion,
for given values of the baryon density, isospin asymmetry
parameter and magnetic field. The last terms in equations 
(\ref{rhps_mag}) and (\ref{rhns_mag}) correspond to the contributions
of the Dirac sea for the scalar densities of proton and neutron.
The magnetized Dirac sea contribution to the nucleon self-energy 
has been calculated by summing over the tadpole diagrams arising 
due to the interaction of the nucleons with the scalar field
$\sigma$ within the Walecka model in the weak magnetic field 
approximation \cite{arghya}. 
Generalizing to include the interactions of the nucleons 
to the strange $\zeta$ and the non-strange isovector $\delta$ 
scalar fields as well, in addition to the interaction with the 
non-strange $\sigma$ field, for the chiral effective model 
used in the present investigation, 
the contribution due to the magnetized Dirac sea 
to the self-energy of the $i$-th nucleon ($i=p,n$) 
is given as
\be
\Sigma_i =\sum_{\alpha=\sigma,\zeta,\delta}
\frac{{g_{\alpha i}}^2}{4\pi^2 m_\alpha^2}
\Big[\frac {(q_i B)^2}{3 m_i^*}+ \{ \Delta_i B)^2 m_i^*
+(|q_i|B)(\Delta_i B)\}\Big \{ \frac{1}{2}+2 \ln \Big (\frac{m_i^*}{m_i}
\Big) \Big\}\Big],
\ee
where, $q_i$ is the charge and $\Delta_i=-\frac{1}{2}\kappa_i\mu_NB$ 
is related to the anomalous magnetic moment of the baryon $i$ 
($p$ and $n$ in the present investigation).

The interactions of the $D(\bar D)$ and $B(\bar B)$ mesons
with the baryons and the scalar mesons are obtained
by genralizing the chiral SU(3) model to the charm and bottom sectors
\cite{amdmeson,amarindamprc,amarvdmesonTprc,amarvepja,DP_AM_bbar}.
For the chiral SU(3) model, the baryon as well as meson
octets can be written in terms of the $3\times 3$ Gell-Mann matrices, 
as $\Phi\sim\lambda_a \phi^a$, $\Phi=B,u,X,V_\mu,A_\mu$. 
However, when the model is generalized to SU(4) to include the charm
hadrons, the meson multiplets (being 15-plet) can be expressed as 
$4\times 4$ Gell-Mann matrices ($\lambda_a$, a=1,...15), 
but the baryon multiplet, being a 20-plet can not be written 
as a square matrix of the same order as meson multiplets. 
When chiral SU(3) model is generalized to the charm 
(and bottom) sectors, the baryons are represented by the tensor, 
$B^{ijk}$, which is antisymmetric in the first two indices.
The baryon-pseudoscalar meson interaction term 
(the Weinberg-Tomozawa term) is then written as
\begin{equation}
{\cal L}_{WT}=-\frac{1}{2}\Big[{\bar B}_{ijk}\gamma^\mu 
\Big ({(\Gamma_\mu)_l}^k B^{ijl} +2 {(\Gamma_\mu)_l} ^j 
B^{ilk}\Big)\Big].
\end{equation}
For the nuclar matter as considered in the present study, 
the relevant entries of the baryon tensor are $B^{121}=-B^{211}$ 
and $B^{122}=-B^{212}$, correspond to $p$ and $n$ respectively.
The masses of the open charm (bottom) mesons are obtained from
the interaction Lagrangian
\begin{equation}
{\cal L}_{int}={\cal L}_{WT}+{\cal L}_{SME}+{\cal L}_{1st range}
+{\cal L}_{d1}+{\cal L}_{d_2},
\label{lag_int}
\end{equation}
where the first term is the Weinberg-Tomozawa term,
${\cal L}_{SME}$ is the scalar exchange term,
and ${\cal L}_{1st range}$, ${\cal L}_{d_1}$
and ${\cal L}_{d_2}$ are the range terms.
The scalar meson exchange term
is obtained from explicit symmetry breaking term given by
(\ref{esb}), with the generalizations:  
$A_p=1/\sqrt{2}m_\pi^2 f_\pi{\mathrm{diag}}(1,1,
\frac{2 m_K^2 f_K}{m_{\pi}^2 f_\pi}-1,
\frac{2 m_D^2 f_D}{m_{\pi}^2 f_\pi}-1)$
and $A_p=1/\sqrt{2}m_\pi^2 f_\pi{\mathrm{diag}}(1,1,
\frac{2 m_K^2 f_K}{m_{\pi}^2 f_\pi}-1,
\frac{2 m_D^2 f_D}{m_{\pi}^2 f_\pi}-1,
\frac{2 m_B^2 f_B}{m_{\pi}^2 f_\pi}-1)$,
and the scalar meson multiplet 
for the SU(4) and SU(5) cases, is given as
$X=diag(\frac{(\sigma-\delta)}{\sqrt 2},
\frac{(\sigma+\delta)}{\sqrt 2},\zeta,\zeta_c)$ and
$X=diag(\frac{(\sigma-\delta)}{\sqrt 2},
\frac{(\sigma+\delta)}{\sqrt 2},\zeta,\zeta_c,\zeta_b)$
respectively, where, $\zeta_c \sim \langle \bar c c \rangle$
and  $\zeta_b \sim \langle \bar b b \rangle$.
The range terms are obtained from the interaction terms
\cite{amarvepja}
\begin{equation}
{\cal L}_{1st \;\; range}=Tr(u_\mu X u^\mu X+X u_\mu u^\mu X),
\end{equation}
\begin{equation}
{\cal L}_{d_1}=\frac{d_1}{4}\big ({\bar B}_{ijk} B^{ijk}
{(u_\mu)_l}^m {(u^\mu)_m}^l\big),
\end{equation}
and,
\begin{equation}
{\cal L}_{d_2}=\frac{d_2}{2}\Big [{\bar B}_{ijk} 
{(u^\mu)_l}^m \Big ({(u^\mu)_m}^k B^{ijl}
+2 {(u^\mu)_m}^j B^{ilk}\Big)\Big]
\end{equation}
In the above equations, $u$ occurring in
in the expressions of $u^\mu$  and $\Gamma^\mu$
given by equations (\ref{u_mu}) and (\ref{gamma_mu}),
is given as, $u=exp(\frac{i}{\sigma_0}\lambda_a \pi^a\gamma^5)$
where, $\lambda_a$, are the $4\times 4$ ($5\times 5$) Gell-Mann matrices 
with $a=1,...15$ ($a=1,...24$) for the generalization 
to the case of SU(4) (SU(5)) model.
The masses of the open charm ($D^\pm,D^0,\bar {D^0}$) and the open bottom
($B^\pm,B^0,\bar {B^0}$) mesons in magnetized (nuclear) matter,
are modified due to their interactions with the nucleons and
the scalar fields \cite{dmeson_mag,bmeson_mag}. The in-medium masses 
are obtained by solving their dispersion relations, which are
obtained from the Fourier tranformations of their equations
of motion. These are given as
\begin{equation}
-\omega^2+ {\vec k}^2 + m_{F(\bar F)}^2
 -\Pi_{F(\bar F)}(\omega, |\vec k|)=0,
\label{dispddbar}
\end{equation}
where $\Pi_{F(\bar F)}$, denotes the self energy of the 
meson $F (\equiv D,B),\;\bar F (\equiv \bar D, \bar B)$ in the medium.
Explicitly, the self energies for the $D$ and $\bar D$ are given as
\cite{dmeson_mag}
\begin{eqnarray}
&&\Pi_D (\omega, |\vec k|)= \frac {1}{4 f_D^2}[3 (\rho_p +\rho_n)
\pm (\rho_p -\rho_n) \big)
] \omega
+\frac {m_D^2}{2 f_D} (\sigma ' +\sqrt 2 {\zeta_c} ' \pm \delta ')
 \nonumber \\
&+& \Big[- \frac {1}{f_D}
(\sigma ' +\sqrt 2 {\zeta_c} ' \pm \delta ')
+\frac {d_1}{2 f_D ^2} (\rho^s_p +\rho^s_n)
+\frac {d_2}{4 f_D ^2} (({\rho^s_p} +{\rho^s_n})
\pm   ({\rho^s_p} -{\rho^s_n}) )\Big ]
(\omega ^2 - {\vec k}^2),
\label{selfd}
\end{eqnarray}
and
\begin{eqnarray}
&&\Pi _{\bar D} (\omega, |\vec k|) =  -\frac {1}{4 f_D^2}[3 (\rho_p +\rho_n)
\pm (\rho_p -\rho_n) ] \omega
+\frac {m_D^2}{2 f_D} (\sigma ' +\sqrt 2 {\zeta_c} ' \pm \delta ')
\nonumber \\ & +& \Big [- \frac {1}{f_D}
(\sigma ' +\sqrt 2 {\zeta_c} ' \pm \delta ')
+\frac {d_1}{2 f_D ^2} (\rho^s_p +\rho^s_n)
+\frac {d_2}{4 f_D ^2} (({\rho^s_p} +{\rho^s _n})
\pm   ({\rho^s_n} -{\rho^s_n})) \Big ]
(\omega ^2 - {\vec k}^2),
\label{selfdbar}
\end{eqnarray}
where the $\pm$ signs refer to the $D^0$ and $D^+$ respectively
in equation (\ref{selfd}) and
to the $\bar {D^0}$ and $D^-$ respectively in equation (\ref{selfdbar}).
For the $B$ meson doublet ($B^+$,$B^0$), and $\bar B$ meson
doublet ( $B^-$, ${\bar B}^0$), the self energies are given by 
\cite{bmeson_mag}
\begin{eqnarray}
&&\Pi _{B} (\omega, |\vec k|) =  -\frac {1}{4 f_B^2} [3 (\rho_p +\rho_n)
\pm (\rho_p -\rho_n) 
] \omega 
+\frac {m_B^2}{2 f_B} (\sigma ' +\sqrt 2 {\zeta_b} ' \pm \delta ')+
\nonumber \\
&& \Big [- \frac {1}{f_B}
(\sigma ' +\sqrt 2 {\zeta_b} ' \pm \delta ')
+\frac {d_1}{2 f_B ^2} (\rho_s ^p +\rho_s ^n)
+\frac {d_2}{4 f_B ^2} (3(\rho^s _p +\rho^s_n)
\pm   (\rho^s_p -\rho^s_n) ) \Big ]
(\omega ^2 - {\vec k}^2),
\label{selfb}
\end{eqnarray}
and
\begin{eqnarray}
&&\Pi _{\bar B} (\omega, |\vec k|)
= \frac {1}{4 f_B^2}[3 (\rho_p +\rho_n)
\pm (\rho_p -\rho_n)] \omega
+\frac {m_B^2}{2 f_B} (\sigma ' +\sqrt 2 {\zeta_b} ' \pm \delta ')+
 \nonumber \\
&& \Big [- \frac {1}{f_B}
(\sigma ' +\sqrt 2 {\zeta_b} ' \pm \delta ')
+\frac {d_1}{2 f_B ^2} (\rho_s ^p +\rho_s ^n
)
+\frac {d_2}{4 f_B ^2} \Big (3(\rho^s_p +\rho^s_n)
\pm   (\rho^s_p -\rho^s_n) \Big ]
(\omega ^2 - {\vec k}^2),
\label{selfbbar}
\end{eqnarray}
where the $\pm$ signs refer to the $B^+$ and $B^0$ respectively
in equation (\ref{selfb}) and to the $B^-$  and $\bar {B^0}$ mesons
respectively in equation (\ref{selfbbar}). The terms in the self-energies
refer to the leading Weinberg-Tomozawa term and the sub-leading terms
(the scalar exchange term and the range terms) in chiral perturbation
expansion. The parameters $d_1$ and $d_2$ are fitted from
the KN scattering lengths \cite{amarvepja}. 
In equations (\ref{selfd})- (\ref{selfbbar}), $\sigma'(=\sigma-\sigma _0)$,
${\zeta_b}'(=\zeta_b-{\zeta_b}_0)$ and  $\delta'(=\delta-\delta_0)$
are the fluctuations of $\sigma$, $\zeta_b$, and $\delta$,
from their vacuum expectation values.

The masses are given as $m_{F(\bar F)}^*=\omega(|\vec k|=0)$,
which depend (through the self energies) on the values 
of the scalar fields ($\sigma$, $\zeta$ and $\delta$)
as well as the number and scalar densities of the nucleons.
In the presence of a magnetic field, the lowest Landau level
(LLL) contributions are taken into account for the charged $D^\pm
(B^\pm)$ mesons. The effective masses of the open charm and bottom
mesons are thus given as
\begin{eqnarray}
&& m_{D^{\pm}}^{eff}=\sqrt {{m_{D^{\pm}}^*}^2+eB},\;\;\;\;\;\;
m_{D^0(\bar {D^0})}^{eff}=m_{D^0(\bar {D^0})}^{*},\,\nonumber
\\
&&m_{B^{\pm}}^{eff}=\sqrt {{m_{B^{\pm}}^*}^2+eB},\;\;\;\;\;\;
m_{B^0(\bar {B^0})}^{eff}=m_{B^0(\bar {B^0})}^{*},\,
\label{mddbar}
\end{eqnarray}
where $m_{F(\bar F)}^*$ is the mass of the open charm (bottom)
meson obtained as solution of the dispersion relation 
given by equation (\ref{dispddbar}).

\begin{figure}
\vskip -3.2in
    \includegraphics[width=1.\textwidth]{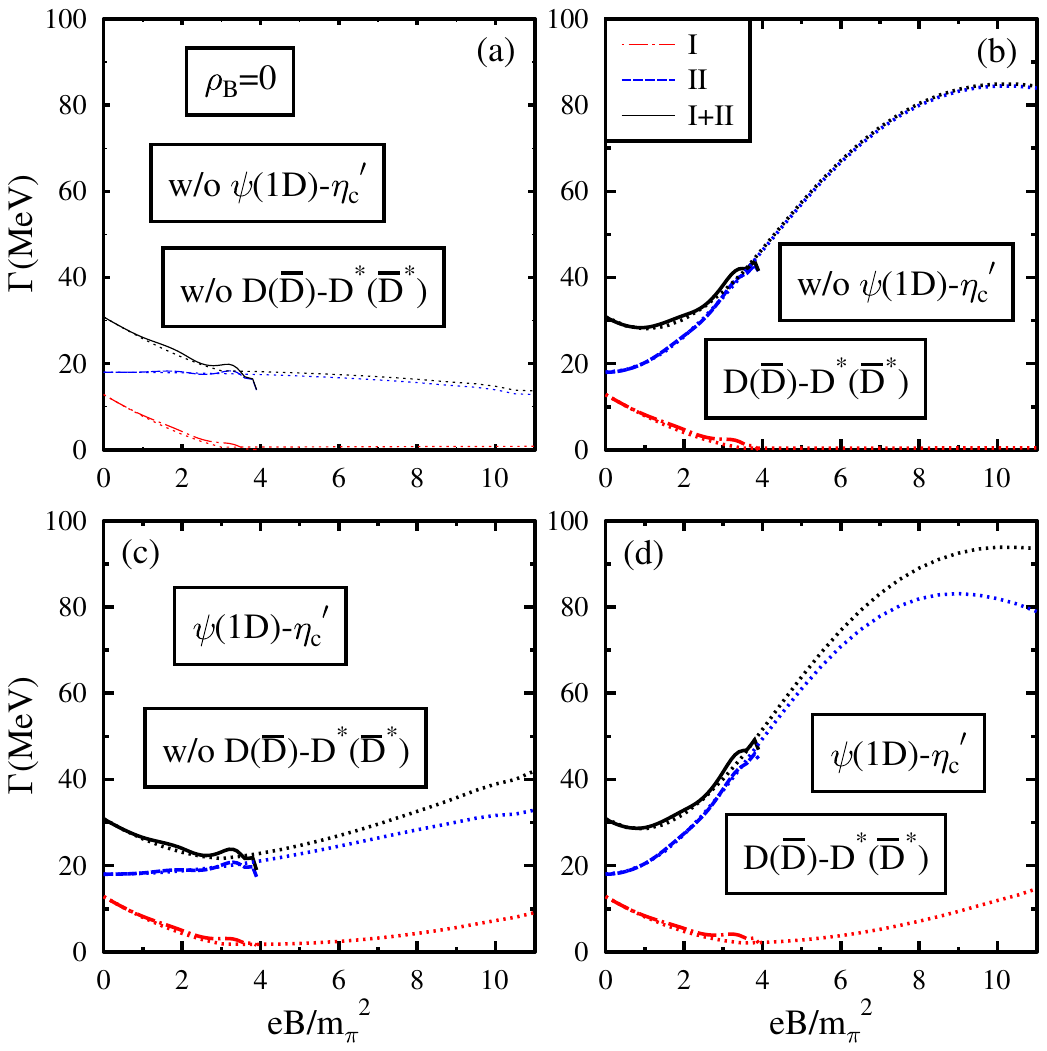}\hfill
\vskip -0.3in
    \caption{Decay widths of (I) $\psi(1D)\rightarrow D^+D^-$,
(II) $\psi(1D)\rightarrow D^0{\bar {D^0}}$, and (III) the sum
of these two channels (I) and (II), as functions of 
$eB/m_\pi^2$, for $\rho_B=0$ with the AMMs of nucleons
taken into account. The effects due to the Dirac sea (DS) 
contributions are included. The effects 
of the $D-D^*$ ($\bar D-\bar {D^*})$ mixing 
on these decay widths are shown in (b) and (d), without and
with the additional effect from $\psi(1D)-\eta_c'$ mixing, 
respectively. The results are compared with the cases when the AMMs
of nucleons are not considered (shown as dotted lines).
}
\label{dwFT_3770_zero_density_MC}
\end{figure}

\begin{figure}
\vskip -3.2in
    \includegraphics[width=1.\textwidth]{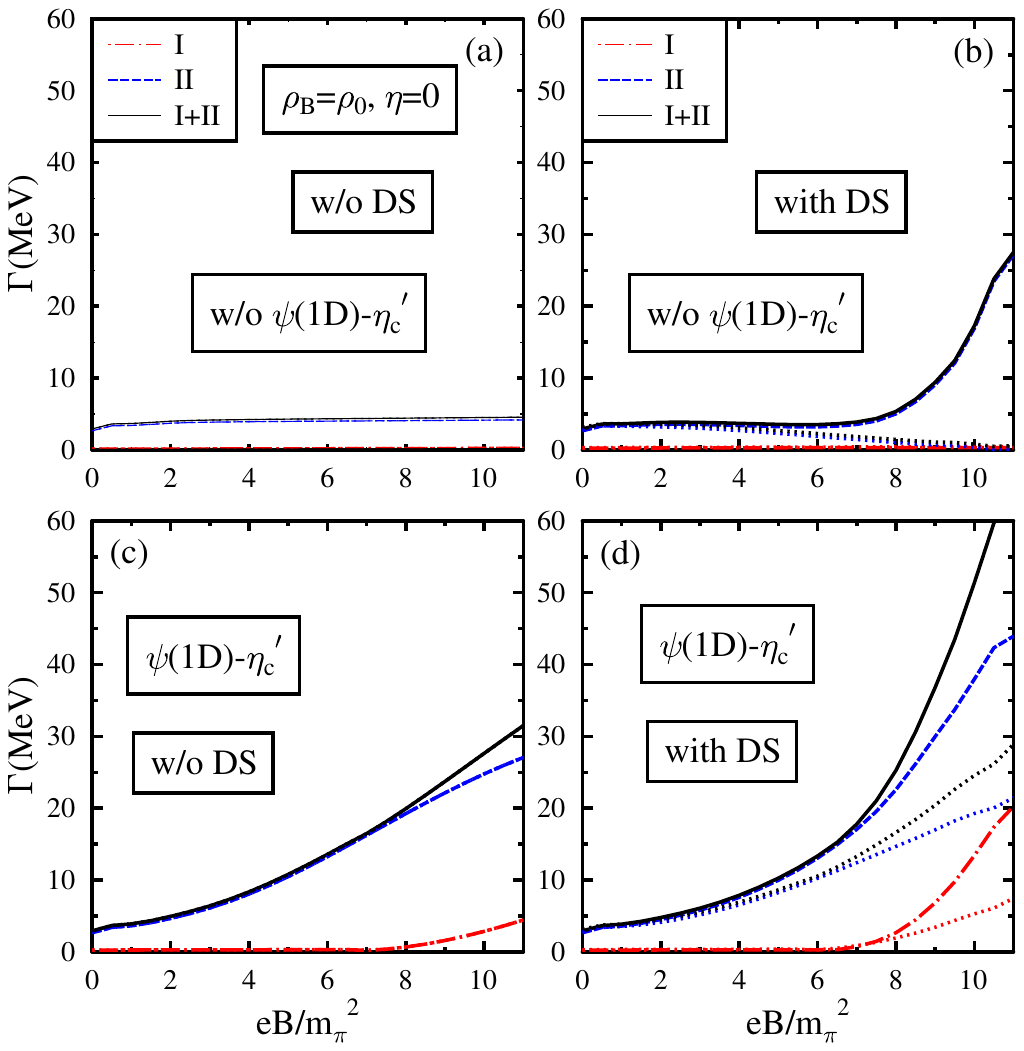}\hfill
\vskip -0.3in
    \caption{Decay widths of (I) $\psi(1D)\rightarrow D^+D^-$,
(II) $\psi(1D)\rightarrow D^0{\bar {D^0}}$, and (III) the sum
of these two channels (I) and (II), as functions of 
$eB/m_\pi^2$, for $\rho_B=\rho_0$ and $\eta=0$ with the AMMs of nucleons
taken into account. The effects due to the Dirac sea (DS) 
contributions are shown in (b) and (d), without and
with the $\psi(1D)-\eta_c'$ mixing, respectively.
The results are compared with the cases when the AMMs
of nucleons are not considered (shown as dotted lines).
}
\label{dwFT_3770_rhb0_eta0_MC}
\end{figure}

\begin{figure}
\vskip -3.2in
    \includegraphics[width=1.\textwidth]{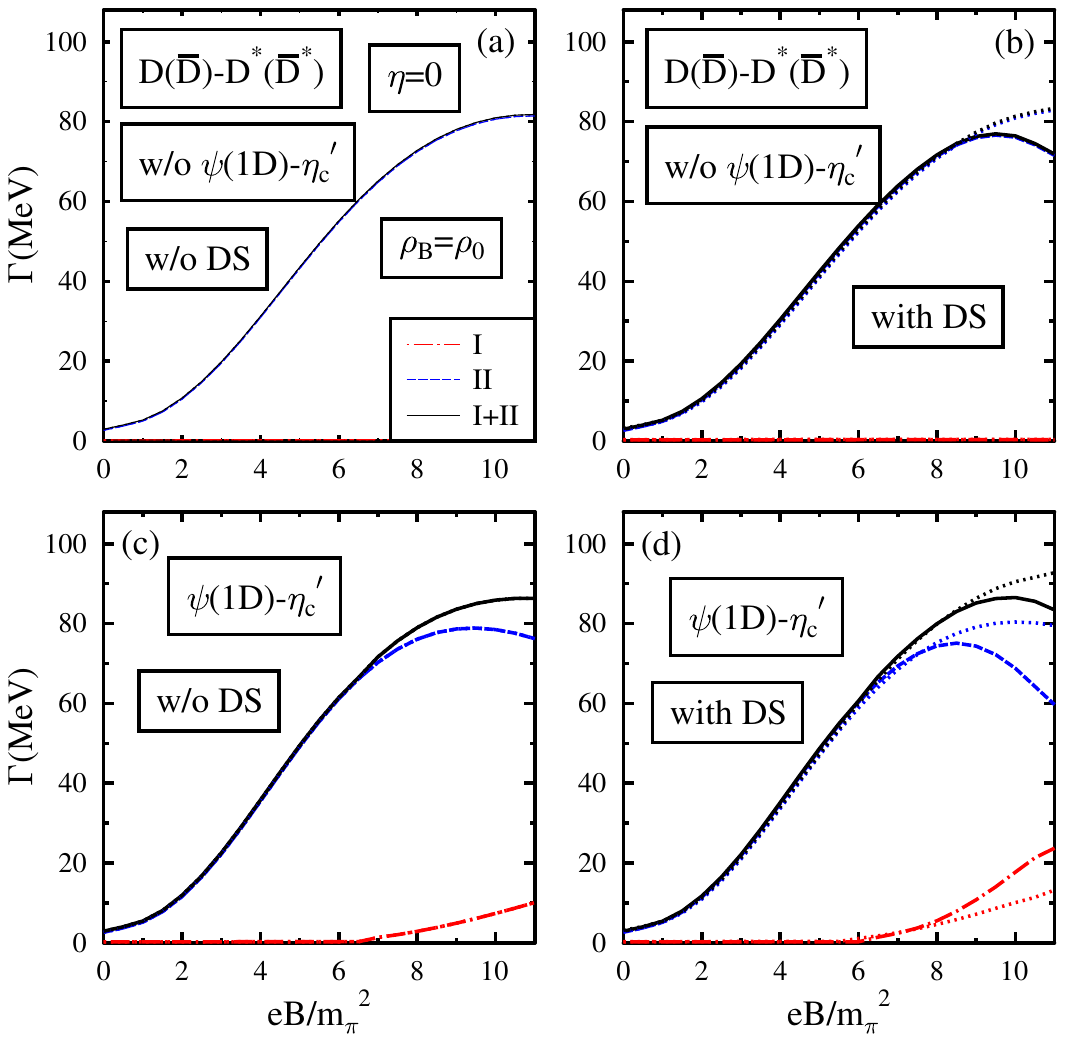}\hfill
\vskip -0.3in
    \caption{Same as Fig. \ref{dwFT_3770_rhb0_eta0_MC},
with additional mass modifications of the open charm mesons
from $D-D^*$ and $\bar D-{\bar {D^*}}$ mixing effects.
\label{dwFT_3770_rhb0_eta0_MC_ds}
}
\end{figure}

%

The mass shift of the heavy quarkonium states 
arises from the medium modification
of the scalar gluon condensate in the leading order 
and is given as \cite{pes1,pes2,voloshin,leeko} 
\begin{eqnarray}
\Delta m_{\Psi(\Upsilon)} &=& \frac{1}{18} \int d{|{\bf k}|}^{2} \langle 
\vert \frac{\partial \psi ({\bf k})}{\partial {\bf k}} \vert^{2} \rangle 
\frac{|{\bf k}|}{|{\bf k}|^{2} 
/ m_{c(b)} + \epsilon} 
\bigg ( 
\left\langle \frac{\alpha_{s}}{\pi} 
G_{\mu\nu}^{a} G^{\mu\nu a}\right\rangle -
\left\langle \frac{\alpha_{s}}{\pi} 
G_{\mu\nu}^{a} G^{\mu\nu a}\right\rangle _{0}
\bigg ),
\label{mass1}
\end{eqnarray}
which, using equation (\ref{chiglu}) gives the mass shift
of the heavy quarkonium state as
\cite{amarvdmesonTprc,amarvepja}
\begin{equation}
\Delta m_{\Psi(\Upsilon)}= \frac{4}{81} (1 - d) \int d |{\bf k}|^{2}
\langle \vert \frac{\partial \psi (\bf k)}{\partial {\bf k}}
\vert^{2} \rangle
\frac{|{\bf k}|}{|{\bf k}|^{2} / m_{c(b)} + \epsilon}
 \left( \chi^{4} - {\chi_0}^{4}\right),
\label{mass_shift}
\end{equation}
where
\begin{equation}
\langle \vert \frac{\partial \psi (\bf k)}{\partial {\bf k}}
\vert^{2} \rangle
=\frac {1}{4\pi}\int
\vert \frac{\partial \psi (\bf k)}{\partial {\bf k}} \vert^{2}
d\Omega.
\end{equation}
In equation (\ref{mass_shift}), $d$ is a parameter introduced
in the scale breaking term in the Lagrangian, $\chi$ and $\chi_0$
are the values of the dilaton field in the magnetized medium 
and in vacuum respectively. 
The wave functions of the quarkonium states,
$\psi(\bf k)$ are assumed to be harmonic oscillator
wave functions, $m_{c(b)}$ is the mass of the charm (bottom) 
quark, $\epsilon=2m_{c(b)}-m_{\psi(\Upsilon)}$
is the binding energy of the charmonium (bottomonium) 
state of mass, $m_{\psi(\Upsilon)}$.
It might be noted here that the leading order mass formula
(given by equation (\ref{mass1}))
was derived using the binding of the heavy quark and antiquark
in the heavy quarkonium state to be Coulombic. This is a good
approximation for the ground state, but not realistic 
for the excited states \cite{leeko}, as the mass shift 
formula contains derivatives of the wave function,
which measure the dipole size of the system.
The wave functions for the charmonium and bottomonium states
are assumed to be harmonic oscillator type, with the strengths
of the potential determined from the rms radii of the 
quarkonium states. 
The mass shifts of the heavy quarkonium states are thus 
obtained from the values of the dilaton field, $\chi$ 
(using equation (\ref{mass_shift})). 
\begin{figure}
\vskip -3.2in
    \includegraphics[width=1.\textwidth]{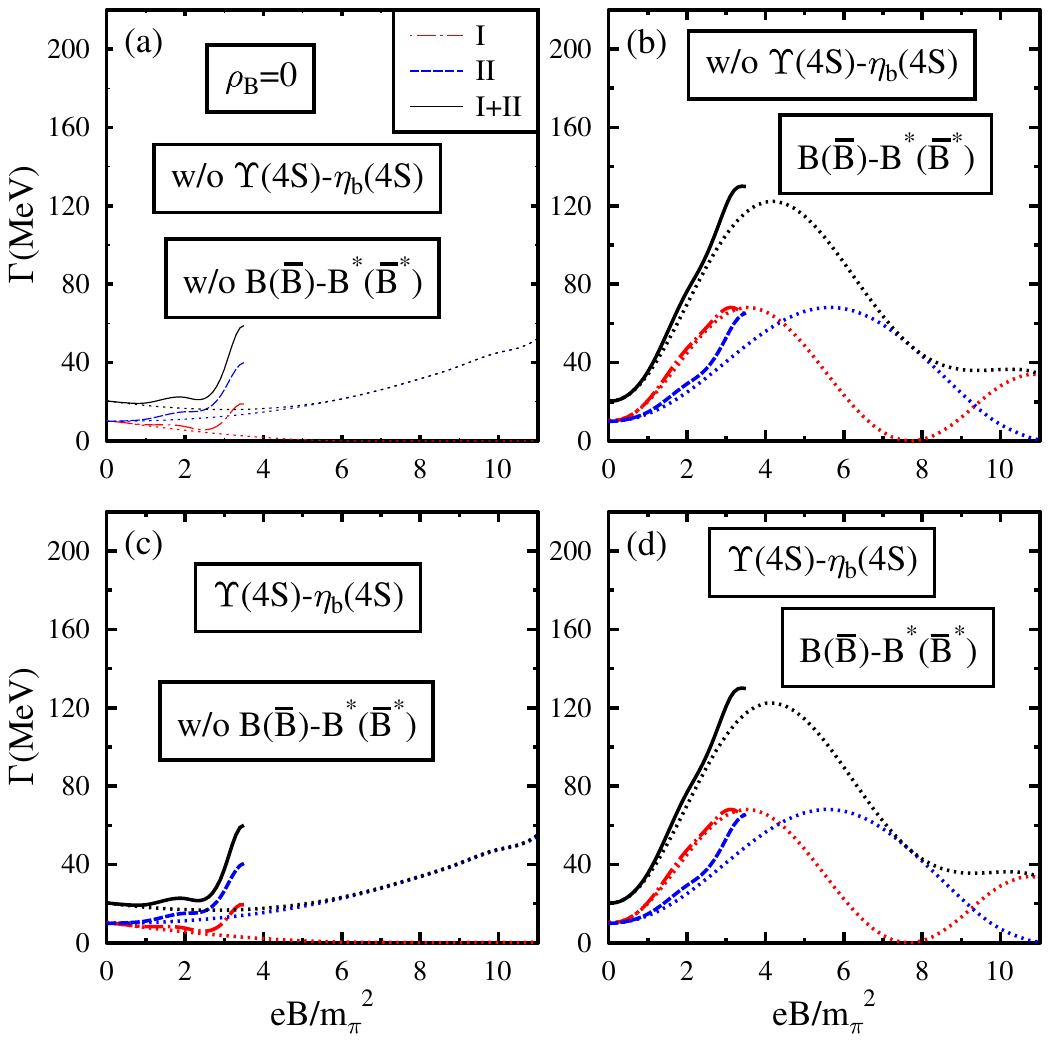}\hfill
\vskip -0.3in
    \caption{Decay widths of (I) $\Upsilon(4S)\rightarrow B^+B^-$,
(II) $\Upsilon(4S)\rightarrow B^0{\bar {B^0}}$, and (III) the sum
of these two channels (I) and (II), as functions of 
$eB/m_\pi^2$, for $\rho_B=0$ with the AMMs of nucleons
taken into account. The effects due to the Dirac sea (DS) 
contributions are included. The effects 
of the $B-B^*$ ($\bar B-\bar {B^*})$ mixing 
on these decay widths are shown in (b) and (d), without and
with the additional effect from $\Upsilon(4S)-\eta_b(4S)$ mixing, 
respectively. The results are compared with the cases when the AMMs
of nucleons are not considered (shown as dotted lines).
}
\label{upsln4s_zero_density_MC}
\end{figure}

\begin{figure}
\vskip -3.2in
    \includegraphics[width=1.\textwidth]{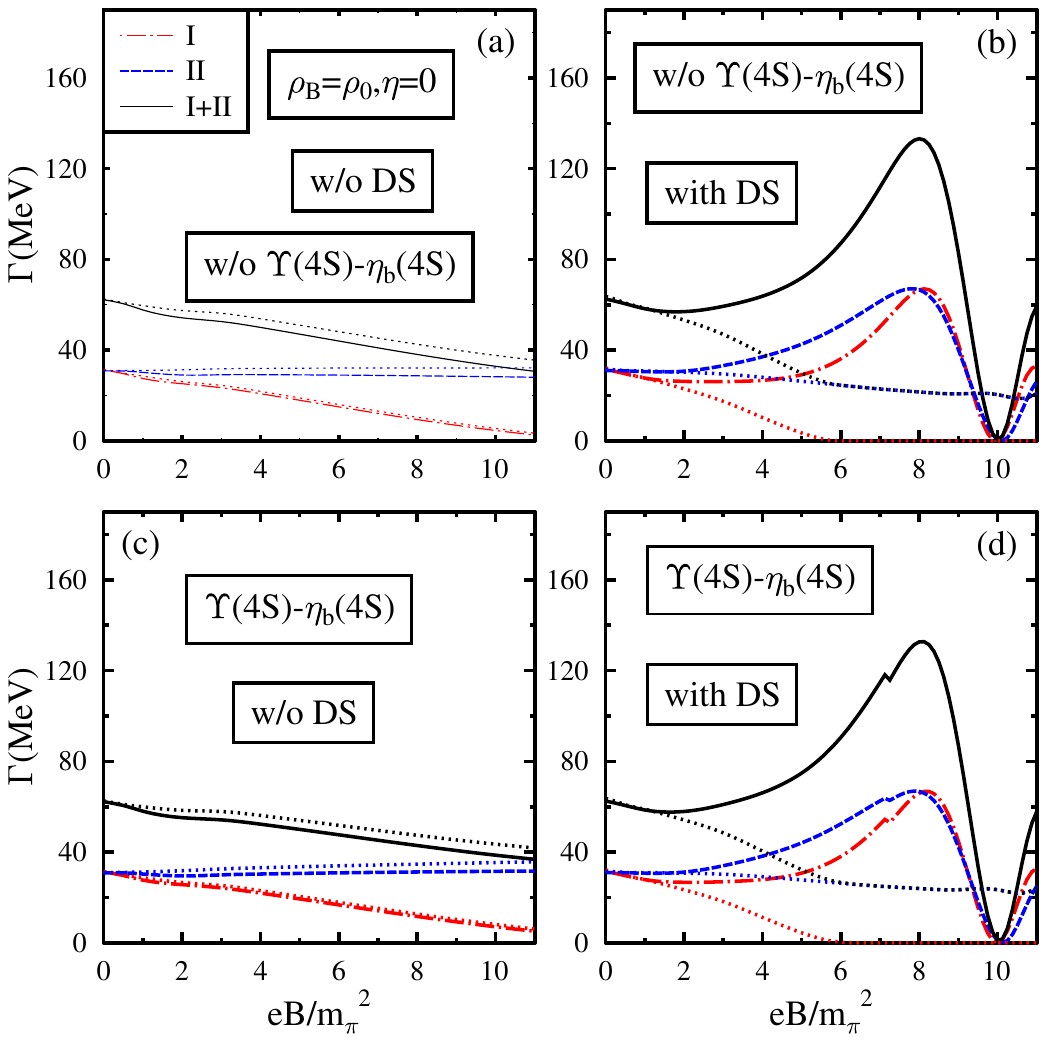}\hfill
\vskip -0.3in
    \caption{Decay widths of (I) $\Upsilon (4S)\rightarrow B^+B^-$,
(II) $\Upsilon(4S)\rightarrow B^0{\bar {B^0}}$, and (III) the sum
of these two channels (I) and (II), as functions of 
$eB/m_\pi^2$, for $\rho_B=\rho_0$ and $\eta=0$ with the AMMs of nucleons
taken into account. The effects due to the Dirac sea (DS) 
contributions are shown in (b) and (d), without and
with the $\Upsilon (4S)-\eta_b(4S)$ mixing, respectively.
The results are compared with the cases when the AMMs
of nucleons are not considered (shown as dotted lines).
}
\label{upsln4s_rhb0_eta0_MC}
\end{figure}

\begin{figure}
\vskip -3.2in
    \includegraphics[width=1.\textwidth]{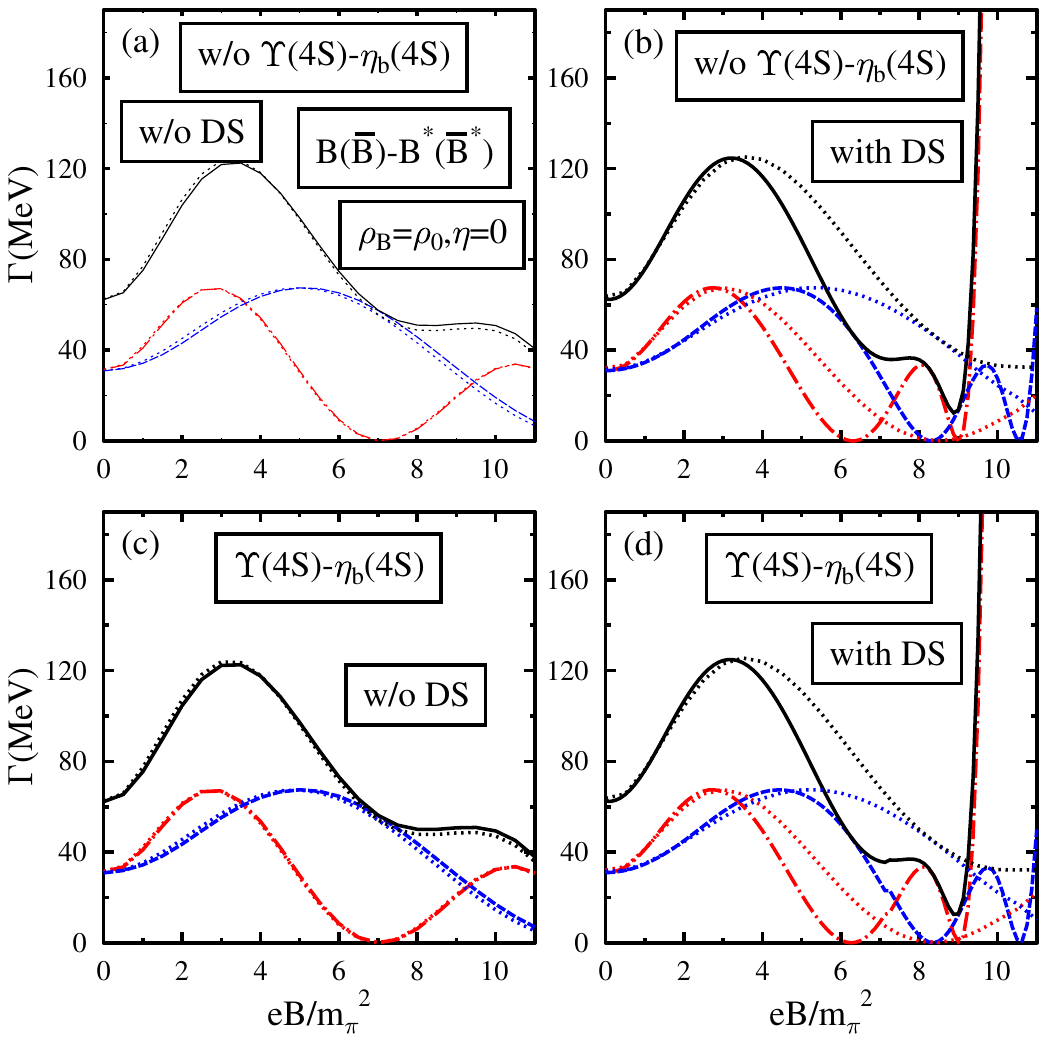}\hfill
\vskip -0.3in
    \caption{Same as Fig. \ref{upsln4s_rhb0_eta0_MC}
with additional mass modifications of the open bottom mesons
from $B-B^*$ and $\bar B-{\bar {B^*}}$ mixing effects.
}
\label{upsln4s_rhb0_eta0_MC_bs}

\end{figure}


\begin{figure}
\vskip -3.2in
    \includegraphics[width=1.\textwidth]{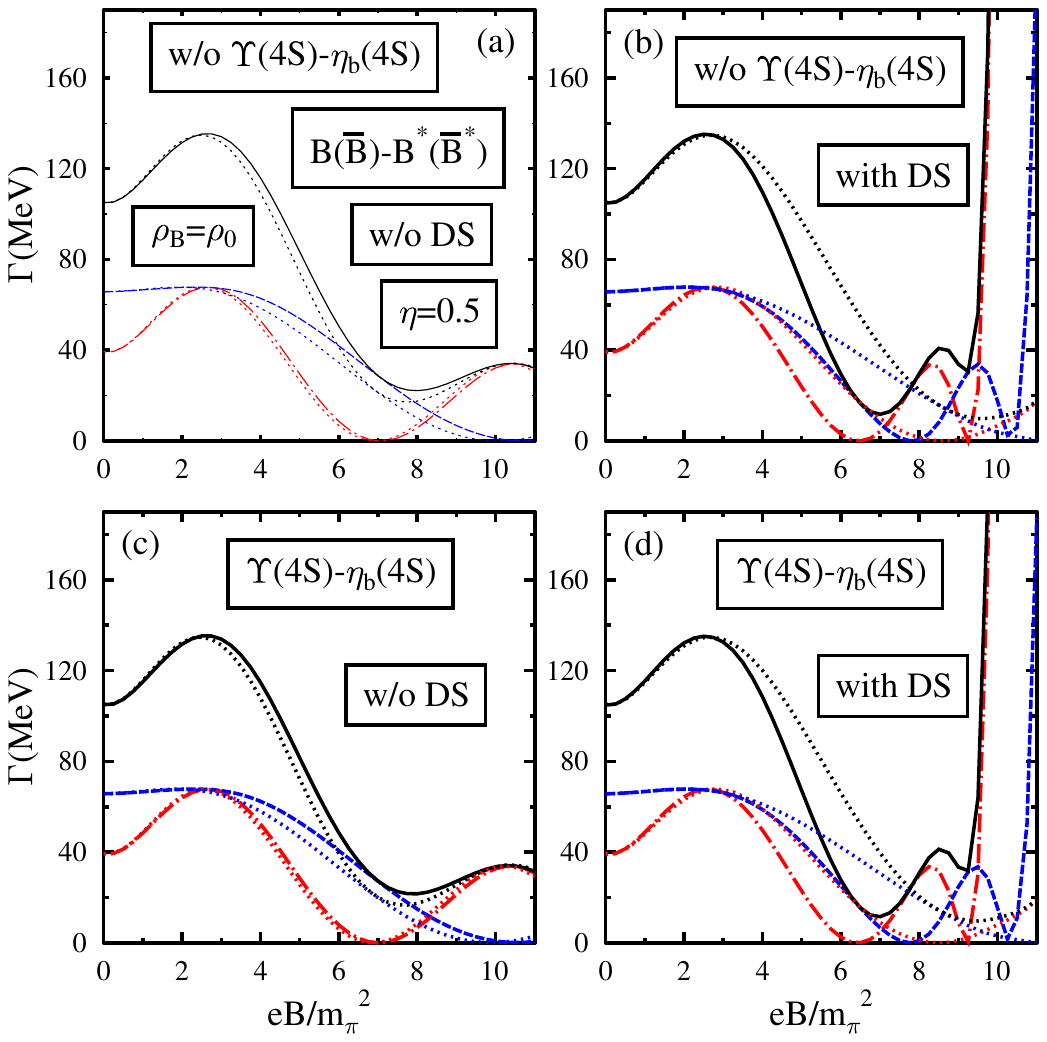}\hfill
\vskip -0.3in
    \caption{same as Fig. \ref{upsln4s_rhb0_eta0_MC_bs},
with $\eta$=0.5.
}
\label{upsln4s_rhb0_eta5_MC_bs}
\end{figure}

The Dirac sea contributions
are included in the scalar densities of the nucleons, which
occur in the equations of motion of the scalar fields,
$\sigma$, $\zeta$ and $\chi$. For given values of 
the baryon density, $\rho_B$, the isospin asymmetry parameter,
$\eta=(\rho_n-\rho_p)/(2\rho_B)$ (with $\rho_n$ and $\rho_p$
as the neutron and proton number densities), the magnetic field, $B$
(chosen to be along z-direction), the fields 
($\sigma$, $\zeta$, $\delta$ and $\chi$) are solved
from their coupled equations of motion.
Within the chiral effective model, 
the masses of the open charm and bottom mesons are given by equations
(\ref{mddbar}), which are obtained from
the solutions of the dispersion relations given by equation 
(\ref{dispddbar}) for $|\vec k|=0$,
with additional Landau level contributions for the charged
mesons.

\subsection{Pseudoscalar-Vector meson (PV) mixing:}

In the presence of a magnetic field, there is mixing between
the spin 0 (pseudoscalar) meson and spin 1 (vector) mesons, which modifies
the masses of these mesons 
\cite{charmonium_mag_QSR,charmonium_mag_lee,Suzuki_Lee_2017,Alford_Strickland_2013,charmdw_mag,open_charm_mag_AM_SPM,strange_AM_SPM,Quarkonia_B_Iwasaki_Oka_Suzuki}.
The PV mixing leads to a drop (rise) in the mass of the
pseudoscalar (longitudinal component of the vector) meson.
The mass modifications have been studied 
using a phenomenological Lagrangian density of the form 
\cite{charmonium_mag_lee,Suzuki_Lee_2017,Quarkonia_B_Iwasaki_Oka_Suzuki} 
\begin{equation}
{\cal L}_{PV\gamma}=\frac{g_{PV}}{m_{av}} e {\tilde F}_{\mu \nu}
(\partial ^\mu P) V^\nu,
\label{PVgamma}
\end{equation}
for the heavy quarkonia \cite{charmonium_mag_QSR,charmonium_mag_lee,Suzuki_Lee_2017,charmdw_mag,upslndw_mag},
the open charm mesons \cite{open_charm_mag_AM_SPM}
and the strange ($K$ and $\bar K$) mesons
\cite{strange_AM_SPM}.
In equation (\ref{PVgamma}), $m_{av}=(m_V+m_P)/2$, 
$m_P$ and $m_V$ are the masses 
for the pseudoscalar and vector charmonium states,
${\tilde F}_{\mu \nu}$ is the dual electromagnetic field.
In equation (\ref{PVgamma}), the coupling parameter $g_{PV}$
is fitted from the observed value of the radiative decay width, 
$\Gamma(V\rightarrow P +\gamma)$.
Assuming the spatial momenta of the heavy quarkonia to be zero,
there is observed to be mixing between the pseudoscalar and the 
longitudinal component of the vector field from their 
equations of motion obtained with the phenomenological
$PV\gamma$ interaction given by equation (\ref{PVgamma}).
The physical masses of the pseudoscalar and the longitudinal component
of the vector mesons including the mixing effects,
obtained by solving their equations of motion, are given as
\cite{charmonium_mag_lee,Suzuki_Lee_2017,Quarkonia_B_Iwasaki_Oka_Suzuki} 
\begin{equation}
m^{(PV)}_{P,V^{||}}=\frac{1}{2} \Bigg ( M_+^2
+\frac{c_{PV}^2}{m_{av}^2} \mp
\sqrt {M_-^4+\frac{2c_{PV}^2 M_+^2}{m_{av}^2}
+\frac{c_{PV}^4}{m_{av}^4}} \Bigg),
\label{mpv_long}
\end{equation}
where $M_+^2=m_P^2+m_V^2$, $M_-^2=m_V^2-m_P^2$ and
$c_{PV}= g_{PV} eB$.
The effective Lagrangian term given by equation
(\ref{PVgamma}) has been observed to lead to the
mass modifications of the longitudinal $J/\psi$ and
$\eta_c$ due to the presence of the magnetic field,
which agree extermely well with
a study of these charmonium states using a
QCD sum rule approach incorporating the mixing effects
\cite{charmonium_mag_QSR,charmonium_mag_lee}.

The PV mixing effects for the open charm mesons 
(due to $D-D^*$ and $\bar D-\bar D^*$ mixings)
\cite{open_charm_mag_AM_SPM}, in addition to the mixing 
of the charmonium states (due to $J/\psi-\eta_c$, $\psi'-\eta_c'$ and
$\psi(3770)-\eta_c'$ mixings)
\cite{charmdw_mag,open_charm_mag_AM_SPM}, 
as calculated using the phenomenological Lagrangian given by 
equation (\ref{PVgamma}) have been observed to lead
to appreciable drop (rise) in the mass of the pseudoscalar
(longitudinal component of the vector) meson. These were
observed to modify the partial decay width of 
$\psi(3770)\rightarrow D\bar D$ 
\cite{charmdw_mag,open_charm_mag_AM_SPM}, with the
modifications being much more dominant due to the 
PV mixing in the open charm 
($D-{D^*}$ and ${\bar D}-{\bar D}^*$)
mesons. 

For the bottom sector, due to lack of data on radiative decays, 
$V\rightarrow P \gamma$, the modifications in the masses 
of the pseudoscalar and vector mesons ($Q_1 {\bar {Q_2}}$ bound states)
due to the PV mixing effects have been estimated from the mixing 
of spin with the external magnetic field
\cite{upslndw_mag}, using the Hamiltonian
\cite{Alford_Strickland_2013,Quarkonia_B_Iwasaki_Oka_Suzuki}.
\begin{equation}
H_{\rm {spin-mixing}}=-{\sum _{i=1}^2} 
{\mbox{\boldmath $\mu$}}_i
\cdot {\bf B},
\label{H_spin_mixing}
\end{equation}
which decribes the interaction of the magnetic 
moments of the quark (antiquark) with the external magnetic field.
In the above, 
${\mbox{\boldmath $\mu$}}_i
=g|e|{q_i} {\bf {S_i}}/(2m_i)$ 
is the magnetic moment of the $i$-th particle, $g$ is the Lande g-factor
(taken to be $2$ $(-2)$ for the quark (antiquark)), $q_i$, $\bf {S_i}$,
$m_i$ are the electric charge (in units of the magnitude of the
electronic charge, $|e|$), spin and mass of the $i$-th particle
\cite{charmonium_mag_lee,Quarkonia_B_Iwasaki_Oka_Suzuki}.
This interaction leads to a drop (increase) of the mass of the 
pseudoscalar (longitudinal component of the vector meson) given as
\cite{Alford_Strickland_2013}
\begin{equation}
{\Delta M}^{PV}= \frac{\Delta E}{2} \Big ( (1+\Delta ^2)^{1/2}-1\Big),
\label{delm_PV}
\end{equation}
where $\Delta=2g|eB|((q_1/m_1)-(q_2/m_2))/\Delta E$,
$\Delta E=m_V-m_P$ is the difference in the masses 
of the pseudoscalar and vector mesons.
It was observed in Ref. \cite{upslndw_mag} that 
the partial decay widths $\Upsilon(4S)
\rightarrow B\bar B$ in presence of an external magnetic field,
calculated using a field theoretical model of composite
hadrons, has significantly larger contributions 
from the PV mixng effects from the open bottom mesons 
($B-B^*$ and $\bar B-\bar {B^*}$ mixings)
as compared to the mixing of the bottomonium states,
$\Upsilon(4S)$ and $\eta_b(4S)$. 
As we shall see later, the inclusion 
of the Dirac sea contributions are observed to lead
to significant modifications to the meson masses,
especially when the AMMs of the nucleons are 
taken into consideration, which, in turn, has 
significant effects on the partial decay widths of
the charmonium (bottomonium) states to the open charm (bottom) mesons.
In the following section, we shall briefly describe
the field theoretical model used to calculate the heavy
quarkonium partial decay widths 
\cite{amspmwg,amspm_upsilon,charmdw_mag,open_charm_mag_AM_SPM,upslndw_mag}.

\section{Partial Decay widths of Charmonium (Bottomonium) state to 
\mbox{\boldmath $D\bar D(B\bar B)$}:} 

In this section, we briefly descirbe the field theoretical 
model of composite hadrons \cite{spm781,spm782,spmdiffscat},
used to study the partial decay widths of the vector 
heavy quarkonium states to open heavy flavour mesons 
in magnetized (nuclear) matter, specifically,
the decay widhts of the charmonium state $\psi(3770)$ 
and the bottomonium state $\Upsilon(4S)$, which are the 
lowest states which decay to $D\bar D$ and $B\bar B$ 
in vacuum. As the matter produced in the non-central 
ultra-relativistic heavy ion collisions (where strong magnetic
fields are created) is dilute,
the quarkonium decay wdiths are studied for vacuum ($\rho_B=0$) 
and for $\rho_B=\rho_0$ in the presence of a magnetic field.
The model used for the calculation of the decay widths
describes the hadrons as comprising of 
quark (and antiquark) constituents. 
The constituent quark field operators of the hadron in motion
are constructed from the constituent quark field operators of
the hadron at rest, by a Lorentz boosting.
Similar to the MIT bag model \cite{MIT_bag}, 
where the quarks (antiquarks) occupy
specific energy levels inside the hadron, it is assumed 
in the present model for the composite hadrons that 
the quark (antiquark) constituents carry fractions
of the mass (energy) of the hadron at rest (in motion) 
\cite{spm781,spm782}.
With explicit constructions of the charmonium (bottomonium)
state and the open charm (bottom) mesons, the decay width
of the heavy quarkonium state to open heavy flavour mesons.
is calculated using the light quark antiquark pair creation
term of the free Dirac Hamiltonian 
for constituent quark field \cite{amspmwg}.
The salient features of the field theoretic model for composite
hadrons are presented in Appendix A.

The relevant part of the quark pair creation term is through the
 $q\bar q (q=u,d)$ creation for decay 
of the charmonium (bottomonium) state, $\Psi$ ($\Upsilon$), 
to the final state, $D\bar D$ ($B\bar B$).  
The pair creation term is given as 
\begin{equation}
{\cal H}_{q^\dagger\tilde q}({\bf x},t=0)
=Q_{q}^{(p)}({\bf x})^\dagger (-i 
\mbox{\boldmath $\alpha$}\cdot
\mbox{\boldmath $\bigtriangledown$} 
+\beta M_q)
{\tilde Q}_q^{(p')}({\bf x}) 
\label{hint}
\end{equation}
where, $M_q$ is the constituent mass of the light quark (antiquark). 
The subscript $q$ of the field operators in equation (\ref{hint})  
refers to the fact that the light antiquark, $\bar q$ and 
light quark, $q$ are the constituents 
of the $D(B)$ and $\bar D(\bar B)$ mesons with momenta ${\bf p}$ and
${\bf p'}$ respectively in the final state of the
decay of the charmonium (bottomonium) state, 
$\Psi(3770)(\Upsilon(4S))$.

Assuming the initial and final state mesons 
to be bound by a harmonic oscillator potential, 
the explict constructions for the vector quarkonium states 
$\psi(3770)$ (corresponding to 1D state) and $\Upsilon (4S)$, 
at rest (with spin projection $m$) 
are given as \cite{spmddbar80,amspmwg,amspm_upsilon}
\begin{equation}
|\psi^m(3770)({\bf 0})\rangle = \frac {1}{4\sqrt {3\pi}} 
{\int {d {\bf k} u_{\psi(1D)}({\bf k})
{c_r} ^i ({\bf k})^\dagger u_r 
\Big ( \bfm\sigma^m -3 (\bfm\sigma \cdot \hat {k})\hat {k}^m \Big )
\tilde {c_s}^i (-{\bf k})v_s|vac\rangle}},
\label{Psi}
\end{equation}
with
\begin{equation}
u_{\psi(1D)}({\bf k})=\Big(\frac{16}{15}\Big)^{1/2} \pi^{-{1}/{4}}
(R_{\psi (1D)}^2)^{7/4} {{\bf k}}^2 \exp \Big (-\frac{1}{2} {R_{\psi(1D)}}^2
{{\bf k}}^2\Big),
\label{upsipp}
\end{equation}
and,
\begin{equation}
|\Upsilon^m(4S)(\vec 0)\rangle = \int {d {\bf k}_1 
b_r^i ({\bf k}_1)^\dagger u_r^\dagger 
u_{\Upsilon(4S)}({\bf k}_1) \sigma^m \tilde {b_s}^i v_s 
(-{\bf k}_1)|vac\rangle},
\label{upsilon}
\end{equation}
with,
\begin{eqnarray}
u_{\Upsilon(4S)}(\bfs k_1)
&=&-\frac{1}{\sqrt{6}}
{\frac{ \sqrt {35}}{4}}
\Bigg({\frac{R_{\Upsilon(4S)}^2}{\pi}}\Bigg)^{3/4}
\left(1-2 R_{\Upsilon(4S)}^2\bfs k_1^2
+\frac{4}{5}R_{\Upsilon(4S)}^4\bfs k_1^4
-\frac{8}{105}R_{\Upsilon(4S)}^6\bfs k_1^6
\right)\nonumber \\
&\times &
\exp\left[-\frac{1}{2}R_{\Upsilon(4S)}^2\bfs k_1^2\right].
\label{u4s}
\end{eqnarray}
In equations (\ref{Psi}) and (\ref{upsilon}), 
${c_r^i}^\dagger ({b_r^i}^\dagger)$
creates a charm (bottom) quark of spin r and color $i$,
${\tilde c}_s^i({\tilde b}_s^i)$ creates 
a charm (bottom) antiquark of spin s and color $i$, 
$S^m\equiv \frac{1}{2}\sigma^m$ gives the spin projection
of the charm (bottom) quark (antiquark),
$u_r$ and $v_s$ are the two component 
spinors for the quark and antiquark. 
The value of the harmonic oscillator strength
for the charmonium state $\psi(3770)$,
is fixed from its rms radius, $r_{rms}$=1 fm 
to be $R_{\psi(3770)}^{-1}$=370 MeV \cite{leeko,amarvepja},
and for the bottomonium state $\Upsilon(4S)$ 
it is fixed from the value of the leptonic decay width
($\Upsilon (4S)\rightarrow e^+e^-$)
of 0.272 keV to be $R_{\Upsilon(4S)}^{-1}$ as 638.6 MeV
\cite{AM_DP_upsilon,amspm_upsilon}.

The states for the open charm and bottom mesons
($F\equiv D,B$, $\bar F \equiv \bar D,\bar B$), 
with finite momenta are constructed in terms of the
constituent quark field operators, obtained 
from the quark field operators of these mesons 
at rest through a Lorentz boosting \cite{spmdiffscat}. 
These are given as
\begin{eqnarray}
&&|F ({\bf p})\rangle  = 
\frac{1}{\sqrt{6}}
\Big (\frac {R_F^2}{\pi} \Big)^{3/4}
\int d{\bf k} 
\exp\Big(-\frac {R_F^2 {\bf k}^2}{2}\Big)
{Q_r}^{i}({\bf k}+\lambda_2 {\bf p})
^\dagger u_r^\dagger 
\tilde {q_s}^{i} 
(-{\bf k} +\lambda_1 {\bf p})v_s
d\bfs k,
\label{F}
\\
&&|{\bar F} ({\bf p}')\rangle 
= \frac{1}{\sqrt{6}}
\Big (\frac {R_F^2}{\pi} \Big)^{3/4}
\int d{\bf k} 
\exp\Big(-\frac {{R_F}^2 {\bf k}^2}{2}\Big)
{q_r}^{i}({\bf k}+\lambda_1 {\bf p}')
^\dagger u_r^\dagger 
\tilde {Q_s}^{i} 
(-{\bf k} +\lambda_2 {\bf p}') v_s
d\bfs k,
\label{Fbar}
\end{eqnarray}
where, for the heavy charm quark, $Q\equiv c$, $q=(d,u)$ 
correspond to the states $(D^+,D^-)$ and $(D^0,\bar {D^0})$ 
respectively, and, for heavy bottom quark $Q\equiv b$, 
$q=(u,d)$ correspond to the open bottom mesons $(B^-,B^+)$ 
and $(\bar B^0,{B^0})$ respectively.
In equations (\ref{F}) amd (\ref{Fbar}),
$\lambda_1$ and $\lambda_2$ are the 
fractions of the mass (energy) of the open charm (bottom) meson
at rest (in motion), carried by the constituent 
light ($q=(d,u)$) antiquark (quark)
and the constituent heavy charm (bottom) quark (antiquark), 
with $\lambda_1 +\lambda_2=1$.
The values of $\lambda_1$ and $\lambda_2$ are calculated 
by assuming the binding energy of the hadron 
as shared by the quark (antquark) to be inversely 
proportional to the quark (antiquark) mass
\cite{spm782,amspmwg,amspm_upsilon}. Taking the constituent
masses of the $u$ and $d$ quarks to be same ($M_u=M_d=M_q$),
the energies of $q(\bar q)$, $(q=u,d)$ and 
$\bar Q (Q)$, with $Q=(c,b)$ in ${\bar F} (F)$ 
meson are then given as
\cite{spm782},
\begin{equation}
\omega_1=M_q+\frac{\mu}{M_q}\times BE  
\;\;{\rm and}\;\;
\omega_2=M_Q+\frac{\mu}{M_Q}\times BE;  
\label{omega12}
\end{equation}
with 
$\mu$ is reduced mass of the light-heavy, $Q\bar q\; (q\bar Q)$ system, 
given as $1/\mu=1/M_Q+1/M_q$ with $M_Q$ and $M_q$ as the constituent
masses of the heavy ($Q$) quark and light ($q$) quark respectively, 
BE is the binding energy, $BE=m_{F(\bar F)}-M_Q-M_q$, and, 
$\lambda_i=\frac{\omega_i}{m_{F(\bar F)}}$, $i=1,2$ are the energies 
carried by the light quark (antiquark) and heavy antiquark (quark).
The motivation for the assumption that the 
contributions from the quark (antiquark) to the binding 
energy of the hadron to be inversely proportional
to the mass of the quark (antiquark) as in equations (\ref{omega12})
is as follows. In fact, in general,
the contributions to the binding energy of the bound state composed
of particles of 1 and 2, with masses $m_1$ and $m_2$, are assumed to be
given as ${\mu}/{m_i}$, $i=1,2$, multiplied by the binding energy 
of the bound state, where, $\mu$ is the reduced mass
of the system, calculated from ${1}/{\mu}={1}/{m_1} +
{1}/{m_2}$.
In other words, the contributions from the particles
to binding energy are inversely proportional to their masses,
and the total binding energy is the sum of the individual 
contributions, i.e., $BE=\big (({\mu}/{m_1})+({\mu}/{m_2})
\big )\times BE =BE$, as it should be.
The reason for making this assumption comes from the example of
hydrogen atom, which is the bound state of the proton and the electron.
As the mass of proton is much larger as compared to the mass of the 
electron, the binding energy contribution from the electron is
$\frac{\mu}{m_e}\times BE \simeq BE$ of hydrogen atom, and the
contribution from the proton is $\frac{\mu}{m_p}\times BE$, 
which is negligible as compared to the total 
binding energy of hydrogen atom, since $m_p >> m_e$. 
With this assumption, the binding 
energies of the heavy-light mesons, e.g., $D$ and $\bar D$ mesons 
as well as for  $B$ and $\bar B$ mesons,
mostly arise from the contribution from the light quark (antiquark).

The decay width of the quarkonium state, $M$, 
for the decay process $M\rightarrow F\bar F$, 
with $(M,F,\bar F)\equiv(\psi(3770),D,\bar D),(\Upsilon(4S),B,\bar B$), 
is calculated from the matrix element of 
the light quark-antiquark pair creation part of the free Dirac Hamiltonian,
between the initial charmonium state and the final state mesons
for the reaction
$M\rightarrow F ({\bf p}) \bar F ({\bf p'})$
as given by
\begin{eqnarray}
\langle F ({\bf p}) | \langle {\bar F} ({\bf p'})|
{\int {{\cal H}_{q^\dagger\tilde q}({\bf x},t=0)d{\bf x}}}
|M^m (\vec 0) \rangle 
= \delta({\bf p}+{\bf p}')A_{M} (|{\bf p}|)p_m,
\label{tfi}
\end{eqnarray}
where,
the expression for $A_M(|{\bf p}|)$ is written in Appendix A.
The decay width is calculated to be
\begin{eqnarray}
\Gamma(M\rightarrow F({\bf p}) {\bar F} (-{\bf p}))
= \gamma_\psi^2\frac{8\pi^2}{3}|{\bf p}|^3
\frac {p^0_{F}(|{\bf p}|) p^0_{\bar F}(|{\bf p}|)}{m_{M}}
A_{M}(|{\bf p}|)^2
\label{gammapsiddbar}
\end{eqnarray}
with $p^0_{F ({\bar F})}(|{\bf p}|)
=\big(m_{F ({\bar F})}^2+|{\bf p}|^2\big)^{{1}/{2}}$, and, 
$|{\bf p}|$, the magnitude of the momentum of the outgoing 
$F(\bar F)$ meson is given as, 
\begin{equation}
|{\bf p}|=\Big (\frac{{m_M}^2}{4}-\frac {{m_F}^2+{m_{\bar F}}^2}{2}
+\frac {({m_F}^2-{m_{\bar F}}^2)^2}{4 {m_M}^2}\Big)^{1/2}.
\label{pd}
\end{equation}
In the above, the masses of the $F(\bar F)$ and heavy 
quarkonium state are the in-medium masses in the magnetized
nuclear matter calculated in the chiral effective model,
with additional contributions from lowest Ladau levels
for the charged open charm (bottom) mesons,
as given by equations (\ref{mddbar}) and (\ref{mass_shift}).
The parameter, $\gamma_M$, in the expression for 
the quarkonium decay width,
is a measure of the coupling strength
for the creation of the light quark antiquark pair,
to produce the $F\bar F$ final state. 
This parameter is adjusted to reproduce the
vacuum decay widths of $\psi(3770)$ to $D^+D^-$ and
$D^0 \bar {D^0}$ \cite{amspmwg} for the charm sector
and $\Upsilon(4S)\rightarrow
B^+B^-$ and  $\Upsilon(4S)\rightarrow
B^0 \bar {B^0}$ \cite{amspm_upsilon} for 
the bottom sector.

When we include the PV mixing effect, the expression
for the decay width is modified to 
\begin{eqnarray}
\Gamma^{PV}(M &\rightarrow & F({\bf p}) {\bar F} (-{\bf p}))
=\gamma_M^2\frac{8\pi^2}{3}
\Bigg [ 
\Bigg(\frac{2}{3} |{\bf p}|^3 
\frac {p^0_F (|{\bf p}|) p^0_{\bar F}(|{\bf p}|)}{m_{M}}
A_{M}(|{\bf p}|)^2 \Bigg)
\nonumber \\
&+&\Bigg(\frac{1}{3} |{\bf p}|^3 
\frac {p^0_F(|{\bf p}|) p^0_{\bar F}(|{\bf p}|)}{m_{M}^{PV}}
A^{M}(|{\bf p}|)^2 \Bigg) \Big({|{\bf p}|\rightarrow |{\bf p|}
(m_{M}= m_{M}^{PV})}\Big)
\Bigg]. 
\label{gammapsiddbar_mix}
\end{eqnarray}
In the above, the first term corresponds to the transverse
polarizations for the quarkonium state, $M$, 
whose masses remain unaffected by the mixing of the 
pseudoscalar and vector charmonium
states. The second term in (\ref{gammapsiddbar_mix})
corresponds to the longitudinal component,
whose mass is modified 
due to the mixing with the pseudoscalar meson 
in the presence of the magnetic field.

\section{Results and Discussions}
We discuss the results obtained due to the effects of 
Dirac sea contributions for the nucleons and the
PV mixing on the decay widths of charmonium state, 
$\psi(3770)\rightarrow D \bar D$ as well as 
$\Upsilon(4S)\rightarrow B\bar B$,
in magnetized isospin asymmetric nuclear matter.
The decay widths are calculated using
a field theoretical model of composite hadrons for 
$\psi(3770)$ and $\Upsilon(4S)$, the lowest 
quarkonium states, which decay to $D\bar D$ and
$\bar B B$ in vacuum. As the created
matter produced in peripheral ultra-relativistic heavy ion
collision experiments, e.g. at RHIC, BNL and at LHC, CERN,
is extremely dilute, we study the effects of the magnetic 
field on the quarkonium partial decay widths at zero density,
and at $\rho_B=\rho_0$, for symmetric as well as asymmetric 
magnetized nuclear matter.
In magnetized nuclear matter, the medium modifications of the quarkonia 
decay widths are obtained from the mass modifications of the initial 
(quarkonia states) and the final (open charm and bottom mesons)
calculated using a chiral effective model
(from equations (\ref{mass_shift}) and (\ref{mddbar})) including the
effects of Dirac sea of the nucleons, with additional LLL contributions
for the charged $D^\pm (B^\pm)$ mesons, which further undergo 
mass modifications due to pseudoscalar meson
-vector meson (PV) mixing in the presence of a magnetic field,
(given by equations (\ref{mpv_long}) and (\ref{delm_PV}) 
for the charm and bottom sectors).
As has already been mentioned, the open charm and bottom meson
masses are obtained from interactions with the nucleons
and scalar mesons ($\sigma$, $\zeta$ and $\delta$)
and mass shifts of the quarkonium states are obtained
from the modifications of a scalar dilaton field, $\chi$,
which mimics the gluon condensates of QCD in the chiral
effective model.
The scalar fields and the dilaton field are solved 
from their coupled equations of motion,
for given values of the baryon density, $\rho_B$, isospin
asymmetry parameter, $\eta$ and the magnetic field, $B$.
In the present study, the AMMs of the nucleons are considered,
which are observed to be important for the mass modifications,
especially, when the Dirac sea effects are taken into 
account.

There is observed to be enhancement of the quark condensates
(calculated from the scalar fields $\sigma$ and $\zeta$
using equation (\ref{qbarq_scalar}))
with increase in the magnetic field, due to Dirac sea
contributions for zero density as well as 
for $\rho_B=\rho_0$ (when AMMs of nucleons are not considered)
both for symmetric ($\eta$=0) and asymmetric (with $\eta$=0.5)
nuclear matter, an effect called magnetic catalysis (MC). 
However, when the AMMs of nucleons are taken into account,
there is observed to be inverse magnetic catalysis (IMC) 
for $\rho_B=\rho_0$, both for symmetric as well as
asymmetric (with $\eta$=0.5) nuclear matter in presence
of a magnetic field. The Dirac sea contributions have 
appreciable effects on the meson masses, and hence
on the decay widths of $\psi(3770)\rightarrow D\bar D$
and $\Upsilon(4S)\rightarrow B\bar B$.
The quarkonium decay widths in magnetized (nuclear) matter were studied 
using a field theoretical model of composite hadrons
\cite{charmdw_mag}, including the effects of the 
mixing of the charmonium (bottomonium) states 
($\psi(3770)-\eta_c(2S)$ ($\Upsilon(4S)-\eta_b(4S)$) mixings)
\cite{charmdw_mag,upslndw_mag}
as well as the PV mixing of the open charm (bottom) mesons
($D(B)-D^*(B^*)$ and $\bar D (\bar B)-\bar {D^*}(\bar {B^*})$ 
mixings) \cite{open_charm_mag_AM_SPM,upslndw_mag}, in addition to the 
Landau level contributions for the charged $D^\pm (B^\pm)$ mesons.
The Dirac sea contributions to the self energies
of the nucleons are observed to lead to important
modifications on the decay widths, which were not considered 
in Refs. \cite{charmdw_mag,open_charm_mag_AM_SPM,upslndw_mag} 
for the mass modifications of the initial and final state 
mesons, hence on the quarkonia decay widths.

Including the effects of the Dirac sea of the nucleons, 
the masses of the open charm \cite{Open_charm_MC},
the bottom meson mesons \cite{Open_bottom_MC},
and the heavy quarkonia states \cite{Heavy_Quarkonia_masses_MC}
have been studied in magnetized (nuclear) matter. 
The inclusion (exclusion) of the AMMs of nucleons 
give rise to the IMC (MC) for $\rho_B=\rho_0$,
which lead to very different behaviours for the masses
of the quarkonium states $\psi(1D)$ and $\Upsilon(4S)$,
with a drop (increase) in the mass, with increase
in the magnetic field, when the PV effects are not 
taken into account \cite{Heavy_Quarkonia_masses_MC}. 
For the open heavy flavour mesons, there is observed 
to be a monotonic increase with magnetic field
when the AMMs are not taken into account,
whereas, there is obserevd to be an intitial increase
followed by a drop in these masses when the magnetic field
is further increased, and the behaviour remains
similar when the PV mixing effects are also taken into
account \cite{Open_charm_MC}. The decay width of the 
quarkonium state $\psi(1D)$ ($\Upsilon(4S)$) (decaying
at rest) to $D\bar D$ ($B\bar B$) depends on the magnitude of
the momentum of the outgoing open heavy flavour mesons,
$|{\bf p}|$, given by equation (\ref{pd}) in terms 
of the in-medium masses of the quarkonium state
and the open heavy flavour mesons. The dependence
of the quarkonium decay width on $|{\bf p}|$ is through a
polynomial term multiplied by an exponential term,
as can be seen from the expression of the decay width 
given by equation (\ref {gammapsiddbar_mix}), in which the 
expression $A^M(|{\bf p}|$ (given by equation (\ref{ap}))
is in the form of an exponential as well as polynomials,
$T_i^M$, whose explicit expressions are written down
in Appendix A.
As we shall see there is observed to be a significant 
difference in the decay width of $\Upsilon(4S)\rightarrow 
B\bar B$, for $\rho_B=\rho_0$, for both symmetric and asymmetric
nuclear matter, when the AMMs are taken 
into account, as compared to when these are ignored.
This is due to the different behavours of the masses 
of the quarkonium and open charm (bottom) mesons,
due to the different behaviours of the scalar fields,
corresponding to (inverse) magnetic catalysis,
in the presence (absence) of the AMMs of the nucleons.
The effects of the Dirac sea contributions 
are seen to be more significant for the
$\Upsilon (4S)\rightarrow B\bar B$, with observation
of nodes at high values of the magnetic field, 
for both the charged and neutral $B\bar B$ 
final state decay widths.

In figure \ref{dwFT_3770_zero_density_MC}, we plot the decay
widths of $\psi(3770)\rightarrow D\bar D$ for $\rho_B=0$ 
including the Dirac sea (DS) contributions for the nucleons
as well as effects from the PV mixing in the presence of
a magnetic field. In the figure \ref{dwFT_3770_zero_density_MC}, 
panel (a) shows the
decay widths of (I) $\psi(3770)\rightarrow D^+D^-$, (II) 
$\psi(3770)\rightarrow D^0{\bar {D^0}}$, and sum of these
sub-channels, in the absence of the PV mixing of the 
charmonium states as well as open charm mesons.
In the absence of the DS contributions, for $\rho_B=0$,
the masses of the charmonium and the neutral open charm  
mesons remain at their vacuum values, but the masses of 
charged mesons $D^\pm$ have positive shifts in the presence
of a magnetic field due to the lowest Landau level (LLL)
contributions. Hence, when the Dirac sea effects are neglected,
the decay width with the $D^0 \bar {D^0}$
final state stays at its vacuum value, whereas the decay width 
of $\psi(3770)\rightarrow D^+D^-$ decreases with increase
in the magnetic field (due to the increase
in the masses of the charged $D^\pm$ mesons),
and, becomes zero at and larger than a certain value of 
magnetic field (when the decay is no longer kinematically possible).
In the presence of the Dirac sea contributions, but when
the PV mixing effects are not taken into account, the masses
of the neutral open charm mesons and charmonium states
are observed to have negligible dependence on the magnetic field 
\cite{Open_charm_MC,Heavy_Quarkonia_masses_MC}. 
However, there is increase in the masses of the charged $D^\pm$ 
mesons due to the lowest Landau level (LLL) contributions, 
which leads to a drop in the decay width for the charged open charm meson 
pair final state in the presence of a magnetic field, whereas, 
the decay width of charmonium to neutral $D\bar D$ is observed 
to drop marginally with increase in the magnetic field, 
in the absence of PV mixing, as can be seen from panel (a) 
in figure \ref{dwFT_3770_zero_density_MC}.
The contributions due to PV mixing have been observed to be 
significant in Ref. \cite{charmdw_mag,open_charm_mag_AM_SPM}.
The Dirac sea contributions
are taken into account using the summation of the tadpole
diagrams, using weak field approximation for the nucleon
propagator \cite{arghya}, and, in the presence of AMMs 
of the nucleons, the solutions do not exist for the scalar 
fields for $eB\ge 4 m_\pi^2$, for $\rho_B=0$ in this approximation. 
The effects of AMMs on the charmonium decay wdiths, 
for this range of magnetic field
where the solutions for the scalar fields and hence the
masses of the open and hidden charm mesons exist,
are observed to be quite small,
as compared to the case when the AMMs are neglected
(shown as the dotted lines). The mixings of the $D-D^*$
and $\bar D -\bar {D^*}$ mesons lead to drop in the
masses of the open charm pseudoscalar mesons, 
and this is observed as a significant enhenacement
of the decay width in the neutral $D\bar D$ channel,
as can be observed in panel (b)
in figure \ref{dwFT_3770_zero_density_MC}.
However, the $D(\bar D)-D^* (\bar D^*)$ mixings are not observed
to affect the decay channel with $D^+D^-$ final state,
the reason for this is due to the fact that the PV effects 
on the masses of the charged
$D^\pm$ mesons become appreciable for higher values 
of magnetic fields ($eB\ge 3 m_\pi^2$) 
\cite{Open_charm_MC}, and, for these values of the
magnetic field, the decay to the charged $D\bar D$ 
is no longer kinematically possible, due to 
the positive Landau level contributions leading to 
increase in the masses of the charged $D^\pm$ mesons.
In the presence of the $\psi(1D)-\eta_c(2S)$ mixing
(which leads to an increase in the mass of the longitudinal 
component of $\psi(1D)$), but without accounting for the mixing 
in the open charm meson sectors, there is observed to
be a rise in the decay widths for both the sub channels 
for high values of magnetic field, as can be seen from panel (c)
in figure \ref{dwFT_3770_zero_density_MC}.
When both the mixings (for the charmonium as well as open charm
mesons) are considered, there is observed to be significant
rise in the charmonium decay width to the neutral $D\bar D$,
as well as, an increase for the charged $D\bar D$ channel
at higher values of the magnetic field, as can be seen 
in panel (d) of figure \ref{dwFT_3770_zero_density_MC}.

The decay widths of $\psi(1D)\rightarrow D\bar D$, along with
the decay widths for the sub-channels (I) $\psi(1D)\rightarrow D^+D^-$ and 
(II) $\psi(1D)\rightarrow D^0 {\bar {D^0}}$, are shown 
for $\rho_B=\rho_0$, accounting for the Dirac sea contributions 
to the scalar densities
of the nucleons as well as with the PV mixing effects
from the charmonium states ($\psi(1D)-\eta_c'$ mxing).
These are shown without and with the PV effects for the open 
charm ($D(\bar D)-D^* (\bar {D^*})$ mixing) mesons, 
for symmetric ($\eta$=0) nuclear matter, 
in figures \ref{dwFT_3770_rhb0_eta0_MC}
and \ref{dwFT_3770_rhb0_eta0_MC_ds} respectively.
When the AMMs of the nucleons are considered,
the Dirac sea contirbutions are observed to 
modify the decay width of charmonium 
to the neutral $D\bar D$ appreciably at high magnetic fields, 
for $\eta=0$ in the absence of PV mixing of open charm mesons.
However, the additonal PV mixing for the charmonium
states ($\psi(1D)-\eta_c'$ mixing), is observed to only
modify the decay widths marginally,
as can be seen from panels (b) and (d) of figure
\ref{dwFT_3770_rhb0_eta0_MC}.
There is observed to be significant rise
in the decay widths when the PV mixing in open charm sectors,
is taken into account, as can be seen
from figure \ref{dwFT_3770_rhb0_eta0_MC_ds}.
The effects of the isospin asymmetry is observed 
to be much less dominant as compared to the effects
due to the Dirac sea contributions and the PV mixing
effects. The AMMs of the nucleons however do play 
an important role, and the Dirac contribution effects 
lead to inverse magnetic catalysis (IMC) when the AMMs 
are considered, whereas, there is observed to be magnetic 
catalysis (MC) when the AMMs are neglected. Due to the 
opposite behaviour of the scalar fields (proportional 
to the light quark condensates), the behaviours of the 
open charm mesons are quite different without and with 
the inclusion of the AMMs of the nucleons, at $\rho_B=\rho_0$,
for symmetric as well as asymmetric nuclear matter
in the presence of a magnetic field. 

In figure \ref{upsln4s_zero_density_MC}, the decay widths
of $\Upsilon(4S) \rightarrow B\bar B$, along with the
sub-channels corresponding to the final states (I) charged and 
(II) neutral $B\bar B$ are shown for $\rho_B=0$, 
taking into account the Dirac sea contributions. 
In panel (a), in the absence of the PV mixings 
for the bottomonium states ($\Upsilon(4S)-\eta_b(4S)$)
as well as for the
open bottom mesons ($B-B^*$ and $\bar B-\bar B^*$),
due to the positive contributions 
to $B^\pm$ masses from Landau levels,
one observes a drop in the width
of the decay to $B^+B^-$ final state with increase
in the magnetic field, which becomes (and remains) 
zero for $eB\ge 5 m_\pi^2$. On the other hand,
the decay width for the neutral $B\bar B$ final state 
shows a steady increase with the magnetic field, 
reaching a value of 45.16 MeV at $eB=10 m_\pi^2$ 
from the vacuum value of around 10 MeV. 
There is observed to be a significant increase 
in the decay widths, more dominant for the $B^+{B^-}$ final
state, due to the PV mixing in the $B(\bar B)-B^*(\bar B^*)$
mesons, as can be seen in panel (b). With further rise
in the magnetic field, there is observed to be a drop 
in the decay widths of both the sub-channels,
reaching zero value (corresponding to the nodes),
for the values of $eB$ of around  7.5 and 11 $m_\pi^2$
for the sub-channels (I) and (II) respectively. 
Similar behaviors of the decay widths are observed 
when the PV mixng in the bottomonium
sector is also taken into account (shown in panel
(d)). However, the PV mixing effects in the
open bottom sector are observed to be much more
appreciable as compared to the PV mixing effect
in the bottomonium sector,
as can be seen in panels (c) and (d) of figure 
\ref{upsln4s_zero_density_MC}.
The observation of the nodes
(vanishing of the decay widths) arises due to
the dependence of the decay widths (given by equation
(\ref{gammapsiddbar_mix})) on the magnitude of the 
momentum of the outgoing $B(\bar B)$ meson (given by equation 
(\ref{pd})) as a polynomial term multiplied by a gaussian
contribution, and the node occurs when the polynomial part
becomes zero. The nodes arise from taking into
consideration the internal structure of the mesons in terms
of the quark and antiquark constituents
\cite{friman,amarvepja,charmdw_mag,upslndw_mag}. 
On the other hand, a phenomenological interaction, 
${\cal L}_{int} \sim \Upsilon ^\mu 
(\bar B(\partial _\mu B) -(\partial_\mu {\bar B})B)$,
without accounting for the internal structure of the mesons, 
leads to the decay widths, which increase monotonically
with increase in $|{\bf p}|$. 

In figure \ref{upsln4s_rhb0_eta0_MC}, the decay widths
are shown for $\rho_B=\rho_0$ for symmetric nuclear
matter, without accounting for the mixings in the
open bottom sector. In panels (a) and (c), these are plotted
for the cases of without and with $\Upsilon(4S)-\eta_b(4S)$
mixing, and, without accounting for the Dirac sea effects.
The decay widths are observed to have significant
contributions with inclusion of Dirac sea effects,
when the AMMs of the nucleons are taken into account,
as can be seen from panels (b) and (d) respectively.
There is observed to be an initial rise and then a
drop and vanishing of the decay widths at around $eB=10 m_\pi^2$, 
for both the sub-channels ((I) $\Upsilon (4S)\rightarrow B^+B^-$
and (II) $\Upsilon (4S)\rightarrow B^0{\bar B}^0$), when the AMMs 
are taken into account. As the magnetic field is further increased, 
there is observed to be increase in these decay widths.
As can be observed in panels (b) and (d) 
in figure \ref{upsln4s_rhb0_eta0_MC}, 
including the Dirac sea effects, when the AMMs of the nucleons
are ignored (shown as dotted lines), 
there is observed to be a drop in the decay width 
(I) $\Upsilon (4S)\rightarrow B^+B^-$
which becomes zero for $eB\sim 6 m_\pi^2$, whereas the decay
width for the neutral $B\bar B$ final state shows a steady, 
but slow decrease with rise in the magnetic field, 
without and with the $\Upsilon(4S)-\eta_b(4S)$ mixing effect
taken into consideration.
The effects on the decay widths of $\Upsilon(4S)\rightarrow B\bar B$ 
from the PV mixing of the bottomonium states 
are observed to be marginal as compared to the effects 
from Dirac sea contributions, as can be seen from figure
\ref{upsln4s_rhb0_eta0_MC}. 
In the presence of Dirac sea effects and AMMs of nucleons, when 
the $\Upsilon(4S)-\eta_b(4S)$ mixing is also taken into account,
there is observed to be a non-smooth behaviour
of the decay widths 
in both charged and neutral $D\bar D$
channels at around $eB\sim 7 m_\pi^2$ (as can be seen 
in panel (d) of figure \ref{upsln4s_rhb0_eta0_MC}). 
This behaviour of the decay widths
(which depend on $|{\bf p}|$) arises from dependence 
of the mass of the bottomonium state, $\Psi(4S)$ 
(hence of $|{\bf p}|$) with the magnetic field,
which is observed to be non-smooth at around this value of $eB$
\cite{Heavy_Quarkonia_masses_MC}.

In figure \ref{upsln4s_rhb0_eta0_MC_bs}, the decay widths
are shown accounting for the $B(\bar B)-B^*(\bar B^*)$ mixings.
In the absence of DS effects, there is observed to be 
appreciable effect due to these mixings 
which are observed to lead to only marginal modifications, 
when the $\Upsilon(4S)-\eta_b(4S)$
mixing is also considered (see panels (b) 
and (d) as compared to panels (a) and (c)). 
There is observed to be a node in the decay width
for the $B^+B^-$ final state
at around $eB\sim 7 m_\pi^2$ in the absence of DS effects,
without and with the PV mixing effects taken into account,
as can be seen from panels (a) and (c).
In the presence of DS effects, the initial rise is followed by
a drop leading to vanishing of the decay width and again an 
increase as the magnetic field is further increased.
The nodes are observed for values of $eB$ around 6.3 (8.2)
and 9 (10.5) $m_\pi^2$, for the charged (neutral) $B\bar B$
final states when the AMMs of the nucleons are taken into
account. The DS effects are observed to be much larger
at higher values of the magnetic fields,
when the AMMs are considered.

In figure \ref{upsln4s_rhb0_eta5_MC_bs}, the decay widths
are shown for $\rho_B=\rho_0$ for asymmetric (with $\eta$=0.5) 
nuclear matter, accounting for the mixings in the bottomonium
as well as the open bottom sector. In panels (a) and (c), 
these are plotted
without DS effects, which are observed to show similar
behaviour as for the symmetric nuclear matter shown in
figure \ref{upsln4s_rhb0_eta0_MC_bs}, however, the values
for $eB=0$, are much higher for the asymmetric nuclear 
matter as compared to the symmetric matter.
In the absence of the DS contributions, for zero magnetic
field \cite{amspm_upsilon}, the different mass modifications 
of the bottomonium state and open bottom mesons
(and hence of values of ${\bf p}|$),
in the asymmetric and symmetric nuclear matter,
lead to the difference in the decay widths of 
$\Upsilon(4S)\rightarrow B\bar B$.
The effects from the PV mixings for the open bottom mesons
are observed to dominate over the effects due to the mixing
in the bottomonium sector, both in the symmetric and
asymmetric nuclear matter.

The magnetic field effects considered on the decay
width of the charmonium (bottomonium) 
state $\psi(1D) \rightarrow D \bar D$ $ (\Upsilon(4S)\rightarrow
B\bar B)$ in the present work, are due to the Dirac sea effects 
of the nucleons, the effects from  
$\psi(1D)-\eta_c'$ $ (\Upsilon(4S)-\eta_b(4S))$,
$D-D^*$ $(B-B^*)$ and $\bar D-\bar {D^*}$ $(\bar B-\bar {B^*})$ 
mixings and Landau level contributions for
the charged, $D^\pm (B^\pm)$ mesons. The Dirac contributions 
are observed to lead to significant modifications 
to the quarkonium decay widths.
The decay of $\Psi(3770)$ to the $D^0\bar {D^0}$
is observed to have much larger contribution
from the Dirac sea effects as compared to
the decay width for the charged $D^+D^-$ final state. 
The effects of the Dirac sea contributions 
are observed to be more significant for the
$\Upsilon (4S)\rightarrow B\bar B$ (as compared
to the decay width of $\psi(1D)\rightarrow D\bar D$).
With Dirac sea effects, there is observed to be a significant 
difference in the decay width of $\Upsilon(4S)\rightarrow 
B\bar B$ in magnetized nuclear matter, for $\rho_B=\rho_0$,
for the cases of ignoring (including) the AMMs 
of the nucleons, when the (inverse) magnetic catalysis 
is observed. The strong magnetic field
created at the early stage should have observable
consequences on the production of the hidden and
open charm mesons arising from ultra-relativistic
heavy ion collision experiments.

\section{summary}
To summarize, we have studied the decay widths
of the charmonium states $\psi(1D)$ to $D\bar D$
and of the upsilon state $\Upsilon(4S)\rightarrow B\bar B$
in magnetized (nuclear) matter, accounting for the
Dirac sea contributions for the self energies
of the nucleons within a chiral effective model.
Th open charm (bottom) mesons are calculated
from their interactions with the nucleons and the
scalar mesons, whereas, the quarkonium masses are calculated
within a chiral effective model from the medium change 
of a scalar dilaton field, which mimics the gluon condensates
of QCD. 
There is observed to be magnetic catalysis effect,
i.e., enhancement of the quark condensates
(given in terms of the scalar fields) with rise in
magnetic field, for $\rho_B=0$, for both the cases
of accounting and ignoring the AMMs of the nucleons.
However, for $\rho_B=\rho_0$, there is observed to be 
inverse magnetic catalysis (IMC) when the AMMs of 
the nucleons are taken into account.
The effects from PV mixing 
($\psi(1D)-\eta_c'$, $D-D^*$ and $\bar D-\bar {D^*}$
mixings for the charm sector and $\Upsilon(4S)-\eta_b(4S)$,
$B-B^*$ and $\bar B-\bar B^*$ mixings for the bottom
sector) in the presnece of the magnetic field
are also taken into account, in addition to the 
Landau contributions for the charged open charm (bottom) 
mesons. The effects of the Dirac sea as well as PV mixings
are observed to be quite significant on the
heavy quarkonium decay widths.
These should have observable consequences
on the production of heavy quarkonium states 
and open heavy flavour mesons, as these are created 
at the early stage of the non-central ultra-relativistic 
heavy ion collision experiments, when the magnetic field
can be still be large.

\acknowledgements
Amruta Mishra acknowledges financial support from Department of 
Science and Technology (DST), Government of India 
(project no. CRG/2018/002226).

\begin{appendices}

\vskip 0.15in
\noindent{\bf {Appendix A: Model for composite hadrons}}

\setcounter{equation}{0}
\renewcommand{\theequation}{A.\arabic{equation}}
\vskip 0.1in

The model describes hadrons comprising of 
quark (and antiquark) constituents.
The field operator for a constituent quark for a hadron at rest
at time, t=0, is written as
\begin {eqnarray}
\psi ({\bf x},t=0)
&=&(2\pi)^{-{3}/{2}}\int \Big [U({\bf k}) u_r q_r ({\bf k})
\exp(i{\bf k} \cdot{\bf x})
+ V({\bf k}) v_s \tilde q_s ({\bf k})
\exp(-i{\bf k} \cdot{\bf x})\Big ] d{\bfs k}\nonumber \\
 & \equiv & Q({\bf x})+\tilde Q({\bf x}),
\label{qx}
\end{eqnarray}
where, $U({\bfs k})$ and $V({\bfs k})$ are given as
\begin{eqnarray}
U({\bfs k})=\left (\begin{array}{c} f(|{{\bf k}}|)\\
{\bfm\sigma}\cdot {\bf k} g(|{{\bf k}}|)\\
\end{array} \right ),\;\;\;
V({\bfs k})=\left (\begin{array}{c} 
{\bfm\sigma}\cdot {\bf k} g(|{{\bf k}}|)\\
f(|{{\bf k}}|)\\
\end{array} \right ),
\label{ukvk}
\end{eqnarray}
The functions $f(|{\bf k}|)$ and $g(|{\bf k}|)$ satisfy the constraint
\cite{spm781},
$f^2+g^2 {\bf k}^2=1$,
as obtained from the equal time anticommutation relation 
for the four-component Dirac field operators. 
These functions, for the case of free Dirac field
of mass $M$, are given as,
\begin{equation}
f(|{\bf k}|)=\left ( \frac{k_0 +M}{2 k_0}\right )^{1/2},\;\;\;\; 
g(|{\bf k}|)=\left ( \frac{1}{2 k_0 (k_0+M)}\right )^{1/2},
\label{fkgk}
\end{equation}
where $k_0=(|{\bf k}|^2+M^2)^{1/2}$. In the above, $M$ is the constituent
quark/antiquark mass.
In equation (\ref{qx}), 
$u_r$ and $v_s$ are the two component spinors for the 
quark and antiquark respectively, satisfying the relations
$u_r^\dagger u_s=v_r^\dagger v_s=\delta_{rs}$.
The operator $q_{r}({\bf k})$ annihilates a quark with spin $r$ 
and momentum ${\bf k}$, whereas, $\tilde q _{s}({\bf k})$
creates an antiquark with spin $s$ and momentum ${\bf k}$,
and these operators satisfy the usual anticommutation relations
\begin{equation}
\{q_{r}({\bf k}),q_{s}({\bf k}')^\dagger\}=
\{\tilde q_{r}({\bf k}),\tilde q_{s}({\bf k}')^\dagger\}=
\delta _{rs} \delta ({\bf k}-{\bf k}').
\end{equation}

The field operator for the constituent quark of hadron with finite 
momentum is obtained by Lorentz boosting the field 
operator of the constituent quark of hadron at rest, which 
requires the time dependence of the quark field operators.
Similar to the MIT bag model \cite{MIT_bag}, 
where the quarks (antiquarks) occupy
specific energy levels inside the hadron, it is assumed 
in the present model for the composite hadrons that 
the quark/antiquark constituents carry  fractions
of the mass (energy) of the hadron at rest (in motion) 
\cite{spm781,spm782}.
The time dependence for the $i$-th quark(antiquark) of a hadron
of mass $m_H$ at rest is given as
\begin{equation}
Q_i(x)=Q_i({\bf x})e^{-i\lambda_i m_H t},\;\;
{\tilde Q}_i(x)={\tilde Q}_i({\bf x})e^{i\lambda_i m_H t},
\label{thadrest}
\end{equation}
where $\lambda_i$ is the fraction of the energy (mass) of the hadron 
carried by the quark (antiquark), with $\sum_i \lambda_i=1$.
For a hadron in motion with four momentum p, 
the field operators for quark annihilation and antiquark creation,
for t=0, are obtained by Lorentz boosting the field operator of the 
hadron at rest, and are given as \cite{spmdiffscat}  
\begin{eqnarray}
Q^{(p)}({\bf x},t) =
\int \frac{d\bfs k}{(2\pi)^{{3}/{2}}}  
S(L(p)) U({\bf k}) 
Q ({\bf k}+\lambda {\bf p})
\exp[{i({\bf k}+\lambda {\bf p})
\cdot{\bf x}-i\lambda p^0 t} ]
\label{qxp}
\end{eqnarray}
and,
\begin{eqnarray}
\tilde Q^{(p)}({\bf x},t) =
\int \frac{d\bfs k}{(2\pi)^{{3}/{2}}} 
S(L(p)) V(-{\bf k}) 
\tilde Q (-{\bf k}+\lambda {\bf p})
\exp[{-i(-{\bf k}+\lambda {\bf p}) \cdot{\bf x}
+i\lambda p^0 t}]. 
\label{tldqxp}
\end{eqnarray}
In the above, $\lambda$ is the fraction of the energy of the hadron, 
carried by the constituent quark (antiquark). 
In equations (\ref{qxp}) and (\ref{tldqxp}),
$L(p)$ is the Lorentz transformation matrix, which yields 
the hadron at finite four-momentum $p$ from the hadron at rest, 
and is given as \cite{spm782}
\begin{equation}
L_{\mu 0}=L_{0 \mu}=\frac {p^\mu}{m_H};\;\;\;\;\;
L_{ij}=\delta_{ij} +\frac {p^i p^j}{m_H (p^0+m_H)},
\label{lp}
\end{equation} 
where, $\mu=0,1,2,3$ and $i=1,2,3$,
and the Lorentz boosting factor $S(L(p))$ is given as 
\begin{equation}
S(L(p))=\Bigg [\frac {(p^0+m_H)}{2m_H}\Bigg ]^{1/2}
+\Bigg [ \frac {1}{2 m_H (p^0+m_H)} \Bigg ]^{1/2} 
{\vec {\alpha}}\cdot {\vec p},
\label{slp}
\end{equation}
where, $\vec \alpha = \left (
\begin{array}{cc}  0 & \vec \sigma \\ \vec \sigma & 0
\end{array}\right)$, are the Dirac matrices.
The Lorentz transformations used to obtain the constitutent quark and 
antiquark operators for hadron at rest to hadron
with momemtum, $p$, as given by equations (\ref{qxp}) 
and (\ref{tldqxp}) have the effect of addition of
the hadron fractional momentum, $\lambda {\bf p}$,
as a translation to the constituent quark (antiquark) 
momentum, ${\bf k}(-{\bf k})$ \cite{spmdiffscat}. 
This is similar to the quasipotential approach, 
where the Lorentz transformation plays the role of a 
translation \cite{quasi_pot_approach}.
Using the composite model picture with 
Lorentz transformations as considered 
in the present work, the various properties of 
hadrons, e.g., charge radii of the proton and pion, 
the nucleon magnetic moments \cite{spm781,spm782}
have been studied.

The pair creation term of the Dirac Hamiltonian density 
\begin{equation}
{\cal H}_{Q^\dagger {\tilde Q}}(x)
=Q(x)^\dagger (-i{\bf {\alpha}}\cdot
{\bf \bigtriangledown} +\beta M)
{\tilde Q}(x) 
\label{pair_creation}
\end{equation}
is used to describe the decay of the heavy charmonium
(bottomonium) state, $M$ at rest to open heavy flavour mesons
$F({\bf p})$ and $\bar F({\bf p'})$.
The operators for the light ($q=u,d$) 
quark and antiquark creation in the above term,
thus belong to different hadrons, $F$ and $\bar F$ with 
4-momenta $p$ and $p'$ respectively. 
The light quark pair creation term of the Hamiltonian density,
is used to describe the decay of a heavy charmonium state
($\bar Q Q$), $Q=b,c$ to $D(B)$ and $\bar D(\bar B)$ states, 
which are bound states of $Q\bar q$ and $\bar Q q$ repsectively,
with light ($u$, $d$) quark antiquark pair creation.
We evaluate the matrix element of 
the quark-antiquark pair creation part of the Hamiltonian,
between the initial charmonium (bottomonium) state 
and the final state, $F\bar F$, $F\equiv (D,B)$,
using explicit constructions for the initial and final state
mesons.
\begin{eqnarray}
\langle F ({\bf p}) | \langle {\bar F} ({\bf p}')|
{ \int {{\cal H}_{q^\dagger\tilde q}({\bf x},t=0)d{\bf x}}}
|{M}_m (\vec 0) \rangle
= \delta({\bf p}+{\bf p}')A^{(M)} (|{\bf p}|)p_m.
\label{tfi_appA}
\end{eqnarray}
With $\langle f | S |i\rangle =\delta_4 (P_f-P_i) M_{fi}$,
we have
$M_{fi}=2\pi (-i A^{M}(|{\bf p}|)p_m$.
For evaluation of the  matrix element of 
the quark-antiquark pair creation part of the Hamiltonian,
between the initial charmonium state and the final state 
$F\bar F$, $F=(D,B)$ state as given by equation (\ref{tfi_appA}),
As the $D(B)$ and $\bar D(\bar B)$ mesons are nonrelativistic, 
we shall assume S(L(p)) and  S(L(p')) to be unity.
We shall also take the approximate forms (with a small momentum
expansion) for the functions $f(|{\bf k}|)$ and $g(|{\bf k}|)$ 
of the field operator as given by 
$g(|{\bf k}|)=1/\left ({2 k_0 (k_0+M)}\right )^{1/2}
\simeq {1}/({2M}),$
and $f(|{\bf k|})=(1-g^2{\bf k}^2)^{1/2} 
\approx 1-((g^2 {\bf k}^2)/2)$ \cite{amspmwg}. 

The expression for the decay width of $M\rightarrow F\bar F$
is obtained as given by equation (\ref{gammapsiddbar}).
The expression for $A^{M}(|\bf p|)$ in the decay width is given as
\begin{eqnarray}
A^{M}(|{\bf p}|) & =& 6c_{M}\exp[(a_{M} {b_{M}}^2
-R_F^2\lambda_2^2){\bf p}^2]
\cdot\Big(\frac{\pi}{a_{M}}\Big)^{{3}/{2}}
\Big[T_0^{M}+T_1^{M}\frac{3}{2a_M}
+T_2^M\frac{15}{4a_{M}^2}\nonumber \\
&+& T_3^M\frac{105}{8a_{M}^3}
+T_4^M\frac{105\times 9}{16a_{M}^4}
\Big],
\label{ap}
\end{eqnarray}
where $a_M$, $b_M$ are given as \cite{amspmwg}
$a_M=\frac{1}{2}R_{M}^2+R_F^2; \;\;\;\; 
b_M=R_F^2\lambda_2/a_M$,
and 
$c_M$, for $M\equiv \psi(3770)$, and $M\equiv \Upsilon(4S)$
are given as
$$c_{\psi (3770)}=\frac{1}{4\sqrt{3\pi}}
\left ({\frac{16}{15}}\right)^{{1}/{2}}
\cdot \pi^{-{{1}/{4}}}
\cdot (R_{\psi(3770)}^2)^
{{7}/{4}}\cdot\frac{1}{6}\cdot
\left(\frac{R_D^2}{\pi}\right)^{{3}/{2}}$$
and,
$$c_{\Upsilon (4S)}=\frac{1}{6\sqrt{6}}
\left(\frac{\sqrt {35}}{4}\right)
\left(\frac{R_{\Upsilon(4S)}^2}{\pi}\right)^{{3}/{4}}
\cdot\left(\frac{R_B^2}{\pi}\right)^{{3}/{2}}.$$
respectively.
In the above expressions, $R_M$ and $R_F$ refer to the 
strengths of the harmonic oscillator wave functions for the
charmonium state, $\psi(3770)$  (bottomonium state $\Upsilon(4S)$), 
and the $F(\bar F)$, $F=D,B$ mesons.

The expressions for $T_{i}^M$ for $M\equiv (\Psi(3770),\Upsilon(4S)$, are given as
\begin{eqnarray}
T_0^{\psi(3770)}&=&2b_{\psi(3770)}^2(1-\lambda_2)\bfs p^2
+2b_{\psi(3770)}^2g^2(\bfs p^2)^2(b_{\psi(3770)}-\lambda_2)((3/2)b_{\psi(3770)}^2
\nonumber \\
&-&(2+\lambda_2)b_{\psi(3770)}+2\lambda_2-(1/2)\lambda_2^2),\nonumber
\end{eqnarray}
\begin{eqnarray}
T_1^{\psi(3770)}&=&g^2\bfs p^2[14{b_{\psi(3770)}}^3-b_{\psi(3770)}^2((32/3)
+(37/3)\lambda_2)\nonumber \\
&+&b_{\psi(3770)}((28/3)\lambda_2-(1/3)\lambda_2^2)],
\nonumber
\end{eqnarray}
\begin{eqnarray}
T_2^{\psi(3770)}&=&g^2[7b_{\psi(3770)}-(2/3)\lambda_2-(4/3)],\nonumber
\end{eqnarray}
\begin{eqnarray}
T_3^{M}&=&0,\;\;\; T_4^M=0.
\end{eqnarray}
\begin{eqnarray}
T^{\Upsilon (4S)}_0
&=& \frac {1}{2} (b_{\Upsilon (4S)}-1) (b_{\Upsilon (4S)}-\lambda_2)
(3 b_{\Upsilon (4S)} +\lambda_2 -4) g^2 |{\bf p}|^2
\nonumber \\
&&\times  
\Bigg (1-2 R_{\Upsilon (4S)}^2 
b_{\Upsilon (4S)}^2 |{\bf p}|^2
+\frac{4}{5} R_{\Upsilon (4S)}^4 
b_{\Upsilon (4S)}^4 {|\bf p|}^4
-\frac{8}{105} R_{\Upsilon (4S)}^6 
b_{\Upsilon (4S)}^6 {|\bf p|}^6
\Bigg)
\nonumber
\end{eqnarray}
\begin{eqnarray}
T^{\Upsilon (4S)}_1
&=&
\frac {g^2}{6} \Big ( 9 (b_{\Upsilon (4S)}-1)
- 2 (3 b_{\Upsilon (4S)} -\lambda_2 -2) \Big )
\nonumber \\
&+& \frac {g^2 {|\bf p|}^2  R_{\Upsilon (4S)}^2}{3}
\Bigg [ (-5 b_{\Upsilon (4S)} +3) (3b_{\Upsilon (4S)} 
+\lambda_2 -4 )(b_{\Upsilon (4S)}-\lambda_2) 
\nonumber \\
&-& 
9 b_{\Upsilon (4S)} ^2 (b_{\Upsilon (4S)}-1) 
+ 2 b_{\Upsilon (4S)} (3 b_{\Upsilon (4S)} -\lambda_2 -2) 
( 3 b_{\Upsilon (4S)} -2)
\Bigg ]
\nonumber \\
&+&\frac {4 g^2 {|\bf p|}^4 R_{\Upsilon (4S)} ^4 b_{\Upsilon (4S)}^2}{15}
\Bigg [ (7b_{\Upsilon (4S)} -5) (3b_{\Upsilon (4S)} 
+\lambda_2 -4 )(b_{\Upsilon (4S)}-\lambda_2) 
\nonumber \\
&+& \frac {9}{2} (b_{\Upsilon (4S)} -1) b_{\Upsilon (4S)} ^2 
-  b_{\Upsilon (4S)} ( 5 b_{\Upsilon (4S)} - 4)
(3b_{\Upsilon (4S)} -\lambda_2 -2 )\Bigg ]
\nonumber \\
&-& 
\frac {8 g^2 {|\bf p|}^6 R_{\Upsilon (4S)} ^6 b_{\Upsilon (4S)}^4}{105}
\Bigg [ \frac {1}{2}(9b_{\Upsilon (4S)} -7) 
(3b_{\Upsilon (4S)} +\lambda_2 -4 )(b_{\Upsilon (4S)}-\lambda_2) 
\nonumber \\
&+&\frac {3}{2} b_{\Upsilon (4S)} ^2 (b_{\Upsilon (4S)}-1) 
-\frac {1}{3} b_{\Upsilon (4S)} (3 b_{\Upsilon (4S)} 
-\lambda_2 -2) (7b_{\Upsilon (4S)} -6)
\Bigg ]
\nonumber
\end{eqnarray}
\begin{eqnarray}
 T^{\Upsilon (4S)}_2
&=&\frac {1}{3} g^2 R_{\Upsilon (4S)}^2 (-9 b_{\Upsilon (4S)} 
-2 \lambda_2 +5)
\nonumber \\
&+& \frac {4}{5} g^2 R_{\Upsilon (4S)} ^4 |{\bf p}|^2
\Bigg [ b_{\Upsilon (4S)} ^2 (7 b_{\Upsilon (4S)} -5) 
\nonumber \\
&+& \frac {1}{6} (3 b_{\Upsilon (4S)} +\lambda_2 -4)
(b_{\Upsilon (4S)} -\lambda_2) (7 b_{\Upsilon (4S)} -3) 
\nonumber \\
 &-& \frac {2}{15}  b_{\Upsilon (4S)} 
(3 b_{\Upsilon (4S)} -\lambda_2 -2)
(21 b_{\Upsilon (4S)} -10) \Bigg ]
\nonumber \\
&+& \frac {4}{5} g^2 R_{\Upsilon (4S)} ^6 |{\bf p}|^4 b_{\Upsilon (4S)} ^ 2
\Bigg [ -\frac {1}{7} b_{\Upsilon (4S)}^2 (9 b_{\Upsilon (4S)} -7)
\nonumber \\
&-&\frac {4}{15} b_{\Upsilon (4S)} (b_{\Upsilon (4S)} -\lambda_2 ) 
(3 b_{\Upsilon (4S)} +\lambda_2 -4)
\nonumber \\
&& -\frac {1}{3} (b_{\Upsilon (4S)}-1) (b_{\Upsilon (4S)} -\lambda_2) 
(3 b_{\Upsilon (4S)} +\lambda_2 -4) 
\nonumber \\
&+ & \frac {2}{105} b_{\Upsilon (4S)} (3b_{\Upsilon (4S)} -\lambda_2 -2) 
(45 b_{\Upsilon (4S)} -28 ) \Bigg ],  
\nonumber 
\end{eqnarray}
\begin{eqnarray}
 T^{\Upsilon (4S)}_3
&=&\frac {2 g^2}{15} R_{\Upsilon (4S)} ^4 (15 b_{\Upsilon (4S)} 
+2 \lambda_2 -5)
\nonumber \\
&+&\frac {4}{5} g^2 R_{\Upsilon (4S)} ^6 |{\bf p}|^2
\Bigg [ -\frac {4}{5} b_{\Upsilon (4S)}^3 
- (b_{\Upsilon (4S)}-1) b_{\Upsilon (4S)} ^2 
\nonumber \\
&-&\frac {2}{21} b_{\Upsilon (4S)} 
(b_{\Upsilon (4S)}-\lambda_2) (3 b_{\Upsilon (4S)} +\lambda_2 -4)
\nonumber \\
&-&\frac {1}{21} (b_{\Upsilon (4S)}-1) (b_{\Upsilon (4S)}-\lambda_2) 
(3 b_{\Upsilon (4S)} +\lambda_2 -4)) 
\nonumber \\
&+& \frac {2}{105} b_{\Upsilon (4S)} (3 b_{\Upsilon (4S)} -\lambda_2 -2)
(27b_{\Upsilon (4S)} -10) \Bigg ] ,
\nonumber 
\end{eqnarray}
\begin{eqnarray}
 T^{\Upsilon (4S)}_4
&=&-\frac {4 g^2 R_{\Upsilon (4S)} ^6 }{35\times 9} 
(21b_{\Upsilon (4S)} + 2 \lambda_2 -5).
\label{fiupsln4s}
\end{eqnarray}
In the expressions for the decay widths of the $\psi(3770)
(\Upsilon (4S)$) state, decaying to $D\bar D (B\bar B)$, 
the parameter, $\gamma_M$ is introduced, which refers to
the production strength of $D\bar D(B\bar B)$ from decay 
of $\Psi(3770) (\Upsilon(4S))$ through light quark 
pair creation. 
This parameter is chosen so as to reproduce the vacuum decay widths for 
the decay channels $M \rightarrow F^+ F^-$ and 
$M \rightarrow F^0 \bar {F^0}$.

\end{appendices}

\begin{thebibliography}{10}
\bibitem{Hosaka}
A. Hosaka, T. Hyodo, K. Sudoh, Y. Yamaguchi, S. Yasui, Prog. Part. Nucl. Phys. \textbf{96}, 88 (2017).

\bibitem{tuchin}
K. Tuchin, Adv. High Energy Phys. \textbf{2013}, 490495 (2013).

\bibitem{eichten_1}
E.Eichten, K. Gottfried, T. Kinoshita, K.D. Lane and T.M. Yan,
Phys. Rev. D {\bf 17}, 3090 (1978).
\bibitem{eichten_2}
E.Eichten, K. Gottfried, T. Kinoshita, K.D. Lane and T.M. Yan,
Phys. Rev. D {\bf 21}, 203 (1980).
\bibitem{satz_1}
L. Kluberg and H. Satz, arXiv:0901.3831 (hep-ph).
\bibitem{satz_2}
F. Karsch, M. T. Mehr, and H. Satz,
Z. Phys. C {\bf 37}, 617 (1988).
\bibitem{satz_3}
A. Bazavov, P. Petreczky, and A. Velytsky,
arXiv: 0904.1748 (hep-ph).
\bibitem{satz_4}
S. Digal, P. Petreczky, and H. Satz,
Phys. Lett. B {\bf 514}, 57 (2001).
\bibitem{satz_5}
A. Mocsy and P. Petreczky,
Phys. Rev. D {\bf 73}, 074007 (2006).
\bibitem{repko}
S.F. Radford and W W. Repko, Phys.Rev D {\bf 75}, 074031 (2007).

\bibitem{Ebert}
D. Ebert, R. N. Faustov, V. O. Galkin, Phys. At. Nuclei {\bf 76},
1554 (2013); {\em ibid}, Eur. Phys. Jour. C {\bf  71}, 1825
(2011).

\bibitem{Bonati_pot_model}
C. Bonati, M. D. Elia, A. Rucci, Phys. Rev. D {\bf 92},
054014 (2015).

\bibitem{Yoshida_Suzuki_heavy_flavour_meson_strong_B}
T. Yoshida and K. Suzuki, Phys. Rev. D {\bf 94}, 074043
(2016).

\bibitem{kimlee}
Sugsik Kim, Su Houng Lee, Nucl. Phys. A {\bf 679}, 517 (2001).

\bibitem{klingl} F. Klingl, S. Kim, S. H. Lee, P. Morath and
W. Weise, Phys. Rev. Lett. {\bf 82}, 3396 (1999).

\bibitem{amarvjpsi_qsr}
Arvind Kumar and Amruta Mishra, Phys. Rev. C {\bf 82}, 045207 (2010).
\bibitem{jpsi_etac_mag}
Pallabi Parui, Ankit Kumar, Sourodeep De, Amruta Mishra,
arXiv: 1811.04622 (nucl-th).
\bibitem{upsilon_etab_mag}
Pallabi Parui, Sourodeep De, Ankit Kumar, and Amruta Mishra, 
arXiv:2104.05471 [hep-ph].
\bibitem{moritalee_1}
K. Morita and S.H. Lee, Phys. Rev. C {\bf 77}, 064904 (2008).
\bibitem{moritalee_2}
S.H. Lee and K. Morita, Phys. Rev. D {\bf 79}, 011501(R) (2009).
\bibitem{moritalee_3}
K. Morita and S.H. Lee, Phys. Rev. C {\bf 85}, 044917 (2012).
\bibitem{moritalee_4}
K. Morita and S.H. Lee, Phys. Rev. Lett {\bf 100}, 022301 (2008).

\bibitem{open_heavy_flavour_qsr_1}
Arata Hayashigaki , Phys. Lett. B {\bf 487}, 96 (2000).
\bibitem{open_heavy_flavour_qsr_2}
T. Hilger, R. Thomas and B. K\"ampfer, Phys. Rev. C {bf 79},
025202 (2009).
\bibitem{open_heavy_flavour_qsr_3}
 T. Hilger, B. K\"ampfer and S. Leupold,
Phys. Rev. C {\bf 84}, 045202 (2011).
\bibitem{open_heavy_flavour_qsr_4}
S. Zschocke, T. Hilger and B. K\"ampfer,
Eur. Phys. J. A {\bf 47} 151 (2011).

\bibitem{Wang_heavy_mesons_1}
Z-G. Wang and Tao Huang, Phys. Rev. C {\bf 84}, 048201 (2011).
\bibitem{Wang_heavy_mesons_2}
Z-G. Wang, Phys. Rev. C {\bf 92}, 065205 (2015).

\bibitem{arvind_heavy_mesons_QSR_1}
Rahul Chhabra and Arvind Kumar,
Eur. Phys. J A {\bf 53}, 105 (2017).

\bibitem{arvind_heavy_mesons_QSR_2}
Rahul Chhabra and Arvind Kumar,
Eur. Phys. J C {\bf 77}, 726 (2017).

\bibitem{arvind_heavy_mesons_QSR_3}
Arvind Kumar and Rahul Chhabra,
Phys. Rev. C {\bf 92}, 035208 (2015).

\bibitem{ltolos} L.Tolos, J. Schaffner-Bielich and A. Mishra,
Phys. Rev. {\bf C 70}, 025203 (2004).
\bibitem{ljhs} L. Tolos, J. Schaffner-Bielich
and H. St\"ocker, Phys. Lett. {\bf B 635}, 85 (2006).
\bibitem{mizutani_1}
T. Mizutani and A. Ramos, Phys. Rev. {\bf C 74},
065201 (2006).
\bibitem{mizutani_2}
 L.Tolos, A. Ramos and T. Mizutani, Phys. Rev. {\bf C 77},
015207 (2008).
\bibitem{HL}J. Hofmann and M.F.M.Lutz,
Nucl. Phys. {\bf A 763}, 90 (2005).

\bibitem{tolos_heavy_mesons_1}
R. Molina, D. Gamermann, E. Oset, and L. Tolos,
Eur. Phys. J A {\bf 42}, 31 (2009).
\bibitem{tolos_heavy_mesons_2}
L. Tolos, R. Molina, D. Gamermann, and E. Oset,
Nucl. Phys. A {\bf 827} 249c (2009).


\bibitem{open_heavy_flavour_qmc_1}
        K. Tsushima, D. H. Lu, A. W. Thomas, K. Saito, and R. H. Landau,
        Phys. Rev. C {\bf 59}, 2824 (1999).
\bibitem{open_heavy_flavour_qmc_2}
        A. Sibirtsev,   K. Tsushima, and A. W. Thomas,
        Eur. Phys. J. A {\bf 6}, 351 (1999).
\bibitem{open_heavy_flavour_qmc_3}
        K. Tsushima and F. C. Khanna, Phys. Lett. B {\bf 552}, 138
        (2003).
\bibitem{qmc_1}
P. A. M. Guichon, Phys. Lett. B {\bf 200}, 235 (1988).
\bibitem{qmc_2}
 K. Saito and A. W. Thomas, Phys. Lett B {\bf 327}, 9 (1994).
\bibitem{qmc_3}
K. Saito, K. Tsushima and A. W. Thomas, Nucl. Phys. A {\bf 609}, 339
(1996).
\bibitem{qmc_4}
 P.K. Panda, A. Mishra, J. M. Eisenberg and W. Greiner,
Phys. Rev. C {\bf 56}, 3134 (1997).
\bibitem{krein_jpsi}
G. Krein, A. W. Thomas and K. Tsushima, Phys. Lett. B {\bf 697},
136 (2011).
\bibitem{krein_17}
G. Krein, A. W. Thomas and K. Tsushima, arXiv: 1706.02688
(hep-ph).


\bibitem {amdmeson}
A. Mishra, E. L. Bratkovskaya, J. Schaffner-Bielich,
S.Schramm and H. St\"ocker, Phys. Rev. {\bf C 69}, 015202 (2004).
\bibitem{amarindamprc}
Amruta Mishra and Arindam Mazumdar, Phys. Rev. C {\bf 79},  024908 (2009).
\bibitem{amarvdmesonTprc}
Arvind Kumar and Amruta Mishra, Phys. Rev. C {\bf 81}, 065204
(2010).
\bibitem{amarvepja}
Arvind Kumar and Amruta Mishra, Eur. Phys. A {\bf 47}, 164
(2011).


\bibitem{DP_AM_Ds}
Divakar Pathak and Amruta Mishra, Adv. High Energy Phys. 2015,
697514 (2015).
\bibitem{DP_AM_bbar}
Divakar Pathak and Amruta Mishra, Phys. Rev. C {\bf 91}, 045206
(2015).
\bibitem{DP_AM_Bs}
Divakar Pathak and Amruta Mishra, Int. J. Mod. Phy. E {\bf 23},
1450073 (2014).
\bibitem{AM_DP_upsilon}
Amruta Mishra and Divakar Pathak, Phys. Rev. C {\bf 90}, 025201
(2014).

\bibitem{pes1} M.E. Peskin, Nucl. Phys. {\bf B156}, 365 (1979).
\bibitem{pes2}
G. Bhanot and M.E. Peskin, Nucl. Phys. {\bf B156}, 391 (1979).
\bibitem{voloshin}
M.B.Voloshin, Nucl. Phys. B154 ,365 (1979).
\bibitem{leeko} Su Houng Lee and Che Ming Ko,
Phys. Rev. C {\bf 67}, 038202 (2003).

\bibitem{Schechter} J. Schechter, Phys. Rev. D {\bf 21}, 3393 (1980).
\bibitem{paper3}
        P. Papazoglou, D. Zschiesche, S. Schramm, J. Schaffner-Bielich,
        H. St\"ocker, and W. Greiner, Phys. Rev. C {\bf 59},  411  (1999).
\bibitem{kristof1}
        A. Mishra, K. Balazs, D. Zschiesche, S. Schramm,
        H. St\"ocker, and W. Greiner,
        Phys. Rev. C {\bf 69}, 024903 (2004).

\bibitem{hartree}
        D. Zschiesche, A. Mishra, S. Schramm, H. St\"ocker and W. Greiner,
        Phys. Rev. C {\bf 70}, 045202 (2004).
\bibitem{kaon_antikaon}
A. Mishra, E. L. Bratkovskaya, J. Schaffner-Bielich, S. Schramm
     and H. St\"ocker, Phys. Rev. C {\bf 70}, 044904 (2004).
\bibitem{isoamss}
A. Mishra and S. Schramm, Phys. Rev. C {\bf 74}, 064904 (2006).
\bibitem{isoamss1}
A. Mishra, S. Schramm and W. Greiner, Phys. Rev. C {\bf 78},
024901 (2008).
\bibitem{isoamss2}
Amruta Mishra, Arvind Kumar, Sambuddha Sanyal, S. Schramm,
Eur. Phys, J. A {\bf 41}, 205  (2009).
\bibitem{pneutronstar}
Amruta Mishra, Arvind Kumar, Sambuddha Sanyal, V. Dexheimer,
S. Schramm, Eur. Phys. J {\bf 45}, 169 (2010).
\bibitem{am_vecmeson_qsr} Amruta Mishra,
Phys. Rev. C {\bf 91} 035201 (2015).
\bibitem{vecqsr_mag}
Amruta Mishra, Ankit Kumar, Pallabi Parui, Sourodeep De,
Phys. Rev. C {\bf 100} 015207 (2019).

\bibitem{charmdecay_mag}
 A. Mishra , A. Jahan CS , S. Kesarwani , H. Raval , S. Kumar, and J.
Meena , Eur. Phys. J. A 55,99 (2019).

\bibitem{friman}
B. Friman, S. H. Lee and T. Song, Phys. Lett, B {\bf 548}, 153
(2002).

\bibitem{3p0_1}
A. Le Yaouanc, L. Oliver, O. Pene and  J. C. Raynal,
Phys. Rev. D {\bf 8}, 2223 (1973).
\bibitem{3p0_2}
A. Le Yaouanc, L. Oliver, O. Pene and  J. C. Raynal,
Phys. Rev. D {\bf 9}, 1415 (1974).
\bibitem{3p0_3}
A. Le Yaouanc, L. Oliver, O. Pene and  J. C. Raynal,
ibid, Phys. Rev. D {\bf 11}, 1272 (1975).
\bibitem{3p0_4}
T. Barnes, F. E. Close, P. R. Page and E. S. Swanson, Phys. Rev. D
{\bf 55}, 4157 (1997).

\bibitem{amspmwg}
Amruta Mishra, S. P. Misra and W. Greiner, Int. J. Mod. Phys.
E {\bf 24}, 155053 (2015).
\bibitem{charmdw_mag}
 Amruta Mishra, S.P. Misra,
Phys. Rev. C {\bf 102}, 045204 (2020).

\bibitem{open_charm_mag_AM_SPM}
Amruta Mishra and S. P. Misra, Int. Jour. Mod. Phys. E {\bf 30},
2150064 (2021).

\bibitem{amspm_upsilon}
Amruta Mishra and S. P. Misra, Phy. Rev. C {\bf 95}, 065206 (2017).

\bibitem{upslndw_mag}
 Amruta Mishra, S.P. Misra,
Int. Jour. Mod. Phys. E \textbf{31} 06, 2250060 (2022).

%
\bibitem{B_mag_QSR} C. S. Machado, R.D. Matheus, S.I. Finazzo and
J. Noronha, Phys. Rev. D {\bf 89}, 074027 (2014).

\bibitem{machado_1} C. S. Machado, F. S. Navarra. E. G. de Oliveira
and J. Noronha, Phys. Rev. D {\bf 88}, 034009 (2013).

\bibitem{Alford_Strickland_2013}
J. Alford and M. Strickland, Phys. Rev. D {\bf 88}, 105017
(2013).

\bibitem{charmonium_mag_QSR} S. Cho, K. Hattori, S. H. Lee, K. Morita
and S. Ozaki, Phys. Rev. Lett. {\bf 113}, 122301 (2014).

\bibitem{charmonium_mag_lee}
S. Cho, K. Hattori, S. H. Lee, K. Morita
and S. Ozaki, Phys. Rev. D {\bf 91}, 045025 (2015).

\bibitem{Suzuki_Lee_2017}
K. Suzuki and S. H. Lee, Phys. Rev. C {\bf 96}, 035203 (2017).

\bibitem{Quarkonia_B_Iwasaki_Oka_Suzuki}
S. Iwasaki, M. Oka, K. Suzuki, Eur. Phys. Jour. A {\bf 57}
(2021) 222.

\bibitem{Gubler_D_mag_QSR} P. Gubler, K. Hattori, S. H. Lee, M. Oka,
S. Ozaki and K. Suzuki, Phys. Rev. D {\bf 93}, 054026 (2016).

\bibitem{kharzeevmc}
D. Kharzeev, K. Landsteiner, A. Schmitt, and H.-U. Yee,
Lect. Notes Phys. \textbf{871}, 1 (2013).

\bibitem{elia}
M. D’Elia, S. Mukherjee, and F. Sanflippo, Phys. Rev. D
\textbf{82}, 051501 (2010).

\bibitem{kharmc1}
D. Kharzeev, Ann. Phys. (N.Y.) \textbf{325}, 205 (2010); K.
Fukushima, M. Ruggieri, and R. Gatto, Phys. Rev. D \textbf{81}, 114031 (2010).

\bibitem{chernodub}
A. J. Mizher, M.N. Chernodub, and E. Fraga, Phys. Rev. D
\textbf{82}, 105016 (2010).
\bibitem{Preis}
F. Preis, A. Rebhan, and A. Schmitt, Lect. Notes Phys. \textbf{871}, 51 (2013).

\bibitem{menezes}
D. P. Menezes, M. Benghi Pinto, S. S. Avancini, and C.
Providencia, Phys. Rev. C \textbf{80}, 065805 (2009); D.P. Menezes, M. Benghi Pinto, S. S. Avancini, A. P. Martinez,
and C. Providencia, Phys. Rev. C \textbf{79}, 035807 (2009).

\bibitem{ammc}
Bhaswar Chatterjee, Hiranmaya Mishra, and Amruta Mishra, 
Phys. Rev. D \textbf{84}, 014016 (2011). 

\bibitem{balicm}
G. S. Bali, F. Bruckmann, G. Endrodi, F. Gruber, and A.
Schaefer, J. High Energy Phys. 04 (2013) 130.

\bibitem{haber}
Alexander Haber, Florian Preis, and Andreas Schmitt, Phys. Rev. D \textbf{90}, 125036 (2014). 

\bibitem{arghya}
Arghya Mukherjee, Snigdha Ghosh, Mahatsab Mandal, Sourav Sarkar, and Pradip Roy, Phys. Rev. D \textbf{98}, 056024 (2018).

\bibitem{kmeson_mag}
Amruta Mishra, Anuj Kumar Singh, Neeraj Singh Rawat, Pratik Aman,
Eur. Phys. Jour. A 55, 107 (2019).

\bibitem{dmeson_mag}
Sushruth Reddy P, Amal Jahan CS, Nikhil Dhale, Amruta Mishra, J. Schaffner-Bielich, Phys. Rev. C \textbf{97}, 065208 (2018).

\bibitem{bmeson_mag}
Nikhil Dhale, Sushruth Reddy P, Amal Jahan CS, Amruta Mishra, Phys. Rev. C \textbf{98}, 015202 (2018).

\bibitem{charmonium_mag}
Amal Jahan CS, Nikhil Dhale, Sushruth Reddy P, Shivam Kesarwani, Amruta Mishra, Phys. Rev. C \textbf{98}, 065202 (2018).

\bibitem{broderick1}
A. Broderick, M. Prakash and J.M.Lattimer, Astrophys. J. \textbf{537}, 351 (2002).

\bibitem{broderick2}
A.E. Broderick, M. Prakash and J. M. Lattimer, Phys. Lett. B \textbf{531}, 167 (2002).

\bibitem{Wei}
F. X. Wei, G. J. Mao, C. M. Ko, L. S. Kisslinger, H. St\"{o}cker, and W. Greiner, J. Phys. G, Nucl. Part. Phys. \textbf{32}, 47 (2006).

\bibitem{strange_AM_SPM}
Amruta Mishra and S. P. Misra, Int. Jour. Mod. Phys. E {\bf 30},
2150014 (2021).

\bibitem{spm781} S. P. Misra, Phys. Rev. D {\bf 18}, 1661 (1978).
\bibitem{spm782} S. P. Misra, Phys. Rev. D {\bf 18}, 1673 (1978).
\bibitem{spmdiffscat} S. P. Misra and L. Maharana, Phys. Rev. D
{\bf 18}, 4103 (1978).

\bibitem{MIT_bag}
A. Chodos, R. L. Jaffe, K. Johnson and C. B. Thorn,
Phys. Rev. D {\bf 10}, 2599 (1974).

\bibitem{spmddbar80} S.P.Misra, K. Biswal and B. K. Parida,
Phys. Rev. {\bf D 21}, 2029 (1980).

\bibitem{Open_charm_MC}
Sourodeep De and Amruta Mishra, arXiv: 2208:09820 (hep-ph).

\bibitem{Open_bottom_MC}
Pallabi Parui, Sourodeep De and Amruta Mishra, arXiv: 2208:10017 (hep-ph).

\bibitem{Heavy_Quarkonia_masses_MC}
Ankit Kumar and Amruta Mishra, arXiv: 2208:14962 (hep-ph).

\bibitem{quasi_pot_approach}
V. G. Kadysievsky, R. M. Mir-Kasimov and N. B. Skachkov,
Nuovo Cimento 55A, 233 (1968).

\end{thebibliography}
\end{document}